\documentclass[10pt]{article}
\usepackage{eurosym}
\usepackage[caption=false]{subfig}
\usepackage{graphicx}
\usepackage{amsmath}
\usepackage{lscape}
\usepackage{amsfonts}
\usepackage{amssymb}
\usepackage{float}

\begin{document}

\title{Solitons supported by singular modulation of the cubic nonlinearity }
\author{Vitaly Lutsky and Boris A. Malomed \\
Department of Physical Electronics, School of Electrical 	   Engineering, \\
Faculty of Engineering, Tel Aviv University, Tel Aviv 69978, Israel\\
 malomed@post.tau.ac.il\\
vitalylutsky@post.tau.ac.il}

\maketitle

\begin{abstract}
A model of the optical media with a spatially structured Kerr nonlinearity
is introduced. The nonlinearity strength is modulated by a set of singular
peaks on top of a self-focusing or defocusing uniform background. The peaks
may include a repulsive or attractive linear potential too. We find that a
pair of mutually symmetric peaks readily gives rise to the spontaneous
symmetry breaking (SSB) of modes pinned to individual peaks, while
antisymmetric pinned modes are always unstable, transforming into robust
spatially odd breathers. Three- and five-peak
structures support symmetric modes, with in-phase or twisted profiles, and
do not give rise to asymmetric states. A stability area is found for the
twisted states pinned to the triple peaks, while the corresponding in-phase
modes are unstable, unless the three modulation peaks are set very close to
each other, covered by a single-peak pinned mode. All patterns pinned to
five peaks are unstable too. Collisions of moving solitons with the
singular-modulation peak are studied too. Slowly moving solitons bounce back
from the peak, while the collisions are quasi-elastic for fast solitons. In
the intermediate case, the soliton is destroyed by the collision. In a
special case, the condition of a resonance of the incident soliton with a
trapped mode supported by the peak leads to capture of the soliton.
\end{abstract}



\section{Introduction and the model}

The great potential offered by solitons for fundamental studies and
development of applications in optics and other areas of physics is commonly
known \cite{Zakharov,Ablowitz,solitons,review,JY}. However, the use of uniform
media as a host material for solitons has its limitations, as they admit the existence of few species of solitary modes. In particular, the integrable one-dimensional
(1D) setting based on the nonlinear Schr\"{o}dinger (NLS)\ equation gives
rise to the single stable class of solutions in the form of fundamental
solitons \cite{Zakharov} (higher-order solutions for breathers are available
too \cite{SY}, but they are not truly stable objects, as their binding
energy is zero).

The variety of stable localized objects may be greatly expanded by using
\textit{pseudopotentials} induced by spatial modulation of the local
nonlinearity strength \cite{Barcelona}. In optics, nonuniform Kerr
nonlinearity can be induced by an inhomogeneous distribution of
nonlinearity-enhancing dopants \cite{Kip}. Similar patterns may be realized
in composite media assembled of different materials \cite{Kominis1,Kominis2,Kominis3,Kominis4}. Still broader flexibility of nonlinearity landscapes is available in Bose-Einstein condensates (BECs), where the landscapes,
underlain by the optical Feshbach resonance \cite{Feshbach,Feshbach2,Feshbach3},
can be ``painted" by rapidly moving laser beams \cite{painting,painting2}.
The magnetic Feshbach resonance can be used too, imposing stationary
nonlinearity patterns by means of magnetic lattices \cite{magnetic,magnetic2,magnetic3}.

In particular, structures in the form of a very narrow region with strong
local cubic nonlinearity, embedded into a linear host setting, are modeled
by the NLS equation with a delta-functional coefficient in front of the
nonlinear term, $g(x)\sim \delta (x)$ \cite{Azbel,Dong,Nir,Yasha,HS} (a discrete
version of this setting is represented by a linear dynamical lattice with an
embedded \ nonlinear site \cite{Tsironis,Tsironis2,Tsironic3,Tsironis4}). In this limit,
the family of solitons pinned to the point-like inhomogeneity (nonlinear
defect) is degenerate, in the sense that their norm does not depend on the
propagation constant/chemical potential, in terms of optics and BEC,
respectively \cite{Nir}. The norm degeneracy is a distinctive feature of
\textit{Townes solitons}, such as those supported by uniform cubic \cite%
{Kuzya,Berge,Fibich} and quintic \cite{Townes-1D} self-focusing in the 2D
and 1D geometry, respectively. The fact that solitons of the Townes type are
always made unstable by the occurrence of the critical collapse in the same
setting makes it necessary to replace the delta-functional nonlinearity
profile by more realistic singular ones, approximated by the cusp,
\begin{equation}
g(x)\sim |x|^{-\alpha },  \label{alpha}
\end{equation}
with $\alpha >0$ \cite{borovkova2012solitons}. The analysis has produced
stable solitons pinned to this nonlinearity peak in the range of $0<\alpha
<1 $. In the limit of $\alpha =1$, the solitons' amplitude vanishes, and
they do not exists at $\alpha >1$, when the singularity is too strong ($\int
g(x)dx$ diverges around $x=0$; actually, solitons cannot be pinned to the
nonlinearity peak with $\alpha \geq 1$ as they are destroyed by the collapse
in that case). It is worthy to note that it was also demonstrated in \cite%
{borovkova2012solitons} that the singular modulation profile with $\alpha <1$
may emulate the spatially uniform attractive cubic nonlinearity in a \textit{%
sub-1D} space, with effective fractal dimension $D=2\left( 1-\alpha \right)
/\left( 2-\alpha \right) <1$.

A natural continuation of the analysis, which is the objective of the
present work, is to study configurations pinned to sets of several
nonlinearity peaks. Such patterns can be created in the experiment by means
of the same techniques that were proposed for building a single peak. In
particular, the possibility of the spontaneous symmetry breaking (SSB) in
patterns attached to a pair of identical peaks, which is known in other
settings \cite{Tsironis,Tsironis2,Tsironic3,Tsironis4}, \cite{Dong,book}, opens a way to
use the double peaks for power-controlled switching in optical circuitry.
Further, sets of several peaks may be considered as a transition to a
periodic lattice of nonlinear defects, which is an object of obvious
interest too \cite{DIRAC1,DIRAC2,Asia,Longhi}.

Thus, we consider the model based on the NLS equation with an $x$-dependent
self-focusing coefficient, $g(x)>0$:%
\begin{equation}
iu_{z}=-\frac{1}{2}u_{xx}+\left[ \sigma -g(x)\right] {\left\vert
u\right\vert }^{2}u+\beta g(x)u.  \label{eq:u_model_1d}
\end{equation}%
This equation is written in terms of optics, with $z$ and $x$ being the
propagation distance and transverse coordinate in a planar waveguide, and $%
\sigma $ representing the nonlinearity coefficient of the uniform background
into which the nonlinearity peaks are embedded. In particular, $\sigma >0$
implies competition of the self-focusing peaks with the self-defocusing
background, cf. work \cite{Kominis4}. The linear-potential term $\sim \beta $
takes into account the fact that material perturbations, which induce the
local modulation of the nonlinearity, may also give rise to similar local
changes of the refractive index. In the application to a cigar-shaped
trapping configuration in BEC with longitudinal coordinate $x$, the
evolutional variable, $z$, in Eq. (\ref{eq:u_model_1d}) must be replaced by
time $t$, and the term $\sim \beta $ accounts for the local linear potential
that may be induced by the same tightly focused laser beams which modify the
local strength of the nonlinearity via the Feshbach resonance.

The singular nonlinearity-modulation patterns considered below are specified
in Table \ref{Table1}, including two, three, or five peaks. For the
completeness' sake, the single peak, which was introduced in \cite%
{borovkova2012solitons}, is included too. In these patterns, $\Delta $
determines the separation between adjacent peaks (for the double peak, the
separation is $2\Delta $), and $\alpha $ is the singularity power in Eq. (%
\ref{alpha}). Coefficients in front of the singular-modulation functions,
which are defined in Table \ref{Table1}, are set to be $1$ by means of
obvious rescaling. Further, the remaining scaling invariance may be used to
fix $|\sigma |=1$ in Eq. (\ref{eq:u_model_1d}), unless we choose $\sigma =0$
(no uniform-nonlinearity background).

It is relevant to stress the difference of the superposition of two singular
peaks, defined as per the second line of Table \ref{Table1}, from the
superposition of regularized $\delta $-functions, represented by Gaussians
with width $a$: while the singular peaks remain discernible at any
separation $2\Delta $ between them, the two peaks in the superposition of
the Gaussians merge into one at $a/\left( 2\Delta \right) >1/\sqrt{2}$ \cite%
{Dong}.
\begin{table}[tbp]
\caption{Different sets of nonlinearity-modulation peaks representing the
coefficient in front of the cubic term in Eq. (\protect\ref{eq:u_model_1d}).}
\label{Table1}\centering
\begin{tabular}[H]{|c|c|}
\hline
1 peak & $g(x) = {\left|x\right|}^{-\alpha}$ \\ \hline
2 peak & $g(x) = \left\vert x-\Delta \right\vert ^{-\alpha }+\left\vert
x+\Delta\right\vert ^{-\alpha }$ \\ \hline
3 peak & $g(x) = \left\vert x-\Delta \right\vert ^{-\alpha }+ {\left|x\right|%
}^{-\alpha}+\left\vert x+\Delta\right\vert ^{-\alpha }$ \\ \hline
5 peak & $g(x) = \left\vert x-2\Delta \right\vert ^{-\alpha } + \left\vert
x-\Delta \right\vert ^{-\alpha }+ {\left|x\right|}^{-\alpha}+\left\vert
x+\Delta\right\vert ^{-\alpha } + \left\vert x+2\Delta\right\vert ^{-\alpha
} $ \\ \hline
\end{tabular}%
\end{table}

The rest of the paper is organized as follows. Methods used in the analysis
are briefly summarized in Section II. These include a quasi-analytical
variational approximation (VA) for the mode pinned to a single peak, and a
numerical scheme elaborated for the identification of stationary patterns
and their stability. The VA presented here generalizes that from \cite%
{borovkova2012solitons}, which did not include the linear-potential
component of the local peak. Basic results for the SSB in patterns pinned to
a pair of peaks are reported in Section III, and findings for the set of
three and five peaks (the latter being a prototype of the lattice of
nonlinear defects) are presented in Section IV. In Section V, a dynamical
situation is considered, \textit{viz}., collisions of free solitons with a
single nonlinearity peak (collisions with defects and interfaces of various
types were studied before \cite{SOLITON1,SOLITON2,Mak,SOLITON6,SOLITON5,SOLITON4,SOLITON3}, but not for the
present one). The paper is concluded by Section VI.

\section{The framework of the analysis}

\subsection{The variational approximation for a single peak}

Stationary solutions to Eq. (\ref{eq:u_model_1d}) with propagation constant $%
k$ are looked for in the usual form, $u\left( x,z\right) =e^{ikz}w(x)$, with
real function $w$ satisfying equation

\begin{equation}
kw-\frac{1}{2}w_{xx}+\left[ \sigma -g({x})\right] {w}^{3}+\beta g({x})w=0.
\label{u_model_1d_single_peak}
\end{equation}%
Numerical solutions of Eq. (\ref{u_model_1d_single_peak}) with zero boundary
conditions at $|x|\rightarrow \infty $ were constructed by means of the
Newton's method, while simulations of the full NLS equation (\ref%
{eq:u_model_1d}) were carried out by means of an explicit Runge-Kutta
algorithm. The stationary solutions are characterized by the total power
(norm),%
\begin{equation}
P=\int_{-\infty }^{+\infty }w^{2}(x)dx.  \label{N}
\end{equation}

The use of the Newton's method is contingent on a possibility to choose a
relevant initial guess. It was provided by attaching to each nonlinearity
peak a localized mode predicted for the isolated defect by the VA. To this
end, the simplest Gaussian ansatz,
\begin{equation}
w(x)=A\exp \left( -\rho x^{2}\right) ,  \label{eq:InitialGuessForm}
\end{equation}%
is substituted in the Lagrangian of Eq. (\ref{u_model_1d_single_peak}) with $%
g(x)$ taken as per the top line in Table \ref{Table1},
\begin{equation}
L=\int_{-\infty }^{+\infty }\left[ \frac{1}{4}\left( \frac{dw}{dx}\right)
^{2}\right] +\left[ \frac{k}{2}w^{2}+\frac{1}{4}\left( \sigma -{\left\vert
x\right\vert }^{-\alpha }\right) w^{4}+\frac{1}{2}\beta \left\vert
x\right\vert ^{-\alpha }w^{2}\right] dx.  \label{L}
\end{equation}%
The calculation of integrals in Eq. (\ref{L}) gives rise to the
corresponding effective Lagrangian,

\begin{gather}
L_{\mathrm{eff}}=\left( 8\sqrt{2}\rho \right) ^{-1}A^{2}\left[ \sqrt{\pi
\rho }\left( \sqrt{2}A^{2}\sigma +4k+2\rho \right) \right.  \notag \\
\left. -2^{\alpha /2}\rho ^{\left( \alpha +1\right) /2}\Gamma \left( \frac{1%
}{2}\left( 1-\alpha \right) \right) \left( 2^{\left( \alpha +1\right)
/2}A^{2}-4\beta \right) \right] ,  \label{Leff}
\end{gather}%
where $\Gamma $ is the Euler's Gamma-function. For given $k$, parameters of
ansatz (\ref{eq:InitialGuessForm}) are determined by the corresponding
Euler-Lagrange equations, $\partial L_{\mathrm{eff}}/\partial \left( A,\rho
\right) =0$.

The use of the VA-predicted profiles for modes pinned to individual peaks as
inputs always provide for convergence of the Newton's method, applied to
settings with one or several peaks, even in cases when (depending on values
of control parameters) the so generated numerical solution might be
conspicuously different from the input.

\subsection{The stability test}

Stability of stationary patterns was checked by means of the linearization
with respect to small perturbations, taking the perturbed solution to Eq. (%
\ref{eq:u_model_1d}) as

\begin{equation}
u\left( x,z\right) =\left[ w(x)+\epsilon _{+}(x)e^{\imath \lambda
z}+\epsilon _{-}(x)e^{-\imath \lambda ^{\ast }z}\right] e^{\imath kz},
\label{eq:general_perturbation}
\end{equation}%
where $\epsilon _{\pm }(x)$ are components of the perturbation eigenmode,
and $\lambda $ is the instability growth rate. Straightforward linearization
reduces the search for eigenvalues $\lambda $ to the calculation of the
spectrum of a linear matrix operator,

\begin{equation}
G=\left(
\begin{array}{cc}
\hat{B} & -C \\
C & -\hat{B}%
\end{array}%
\right) ,  \label{G}
\end{equation}%
where $C\equiv \left[ \sigma -g(x)\right] w^{2}(x)$, and the Sturm-Liouville
operator is

\begin{equation}
\hat{B}=-k+\frac{1}{2}\frac{d^{2}}{dx^{2}}-2\left[ \sigma -g(x)\right]
|w(x)|^{2}-\beta g(x)  \label{A_in_characteristic_matrix}
\end{equation}%
[in fact, the stability analysis is presented below for solutions with real $%
w(x)$]. Dealing with the conservative systems, stable stationary solutions
are identified as those for which all eigenvalues $\lambda $ have zero real
parts. The so predicted stability was checked in direct simulations of the
evolution of perturbed solutions of Eq. (\ref{eq:u_model_1d}).

\section{\textbf{The system with a symmetric pair of singular-modulation
peaks }}

As said above, the basic issue for the setting with the double peak, which
corresponds to Eq. (\ref{eq:u_model_1d}) with $g(x)$ taken as per the second
line in Table \ref{Table1}, is the possibility of the SSB between solitons
pinned to the two peaks, with separation $2\Delta $ between them. The
asymmetry is characterized by the normalized intensity-contrast parameter,%
\begin{equation}
\Theta \equiv \frac{w^{2}(x=+\Delta )-w^{2}(x=-\Delta )}{w^{2}(x=+\Delta
)+w^{2}(x=-\Delta )}.  \label{Theta}
\end{equation}

\begin{figure*}[tbp]
\centering\subfloat[]{
	\includegraphics[width=0.32\textwidth]{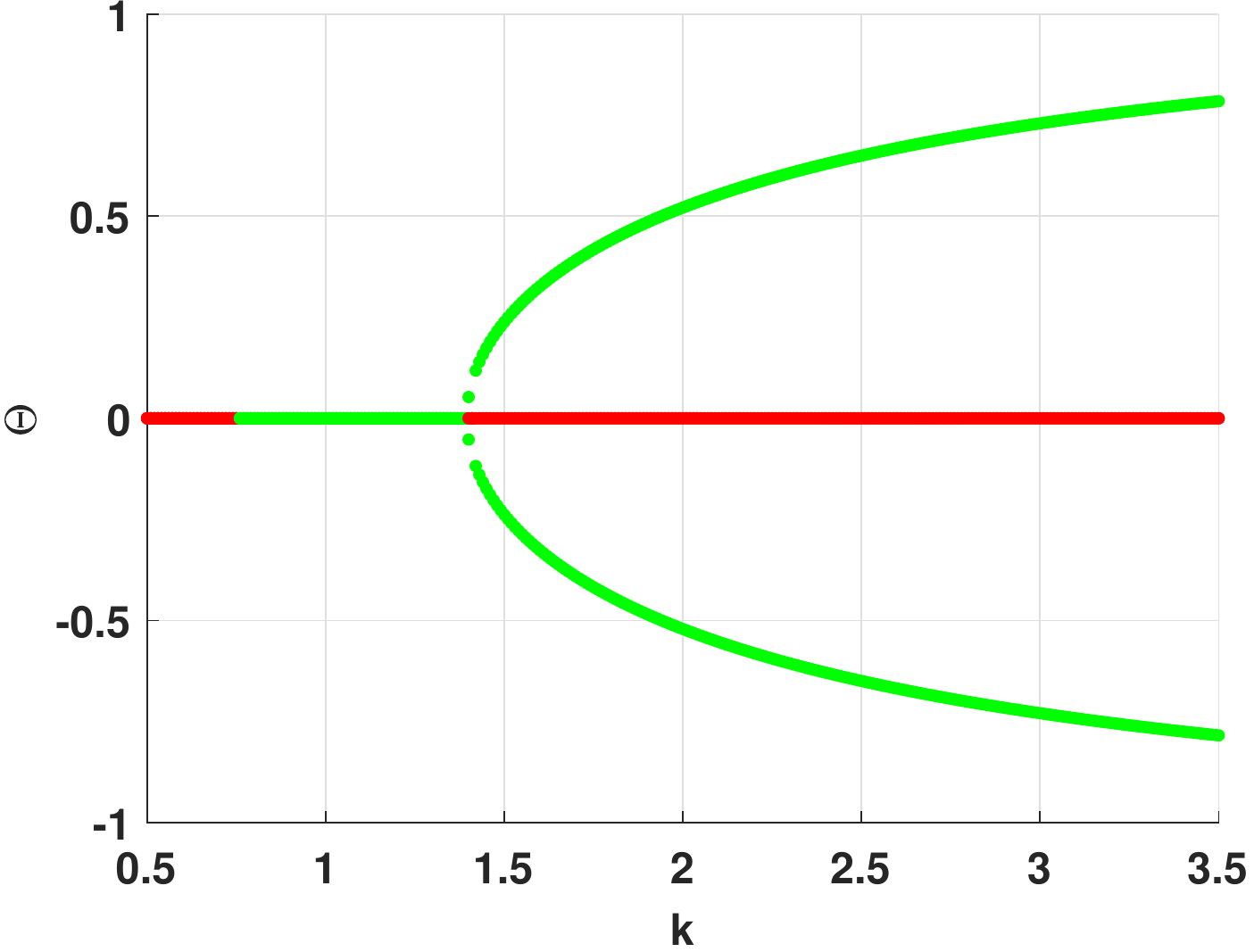}
	}\subfloat[]{
	\includegraphics[width=0.32\textwidth]{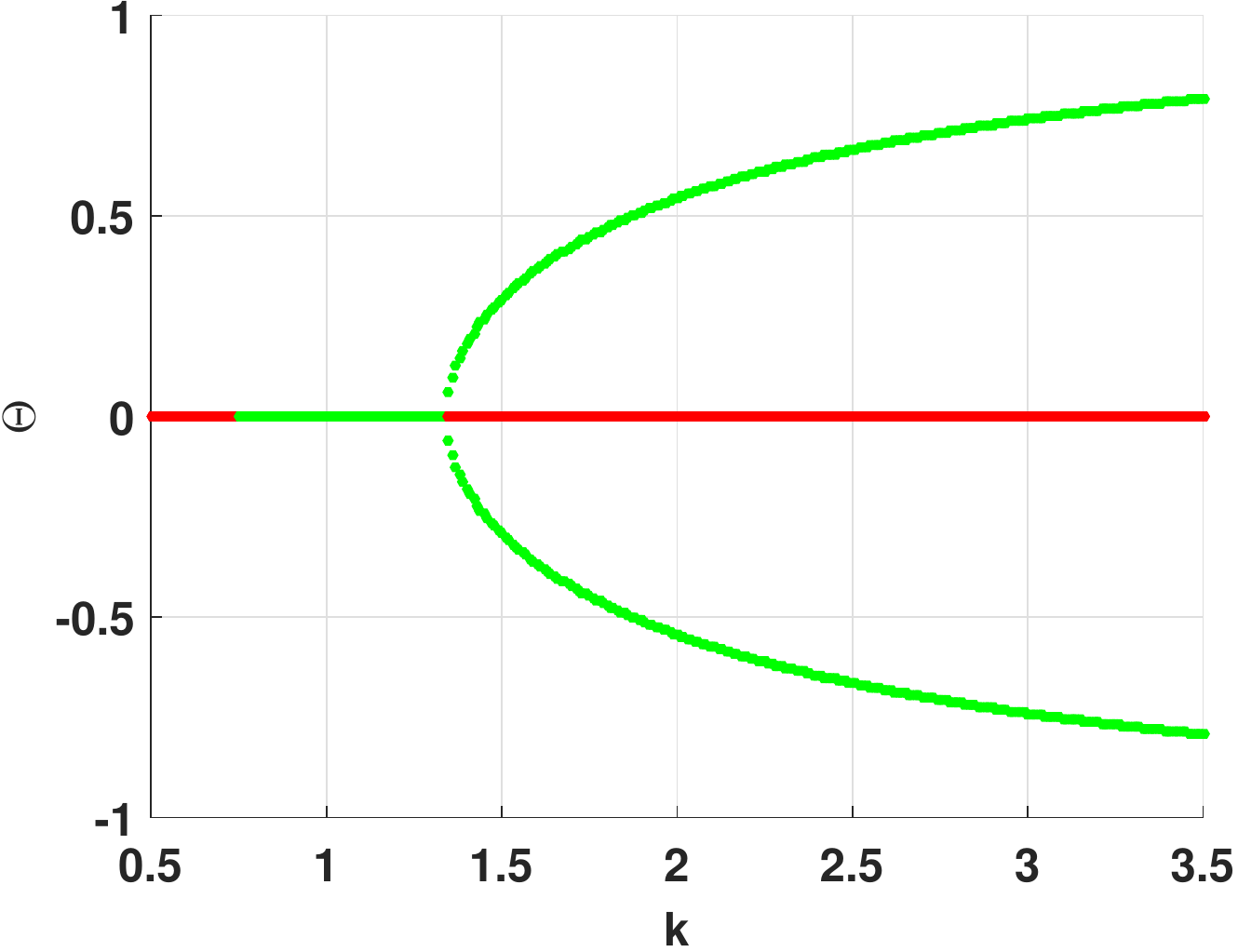}
	}\subfloat[]{
	\includegraphics[width=0.32\textwidth]{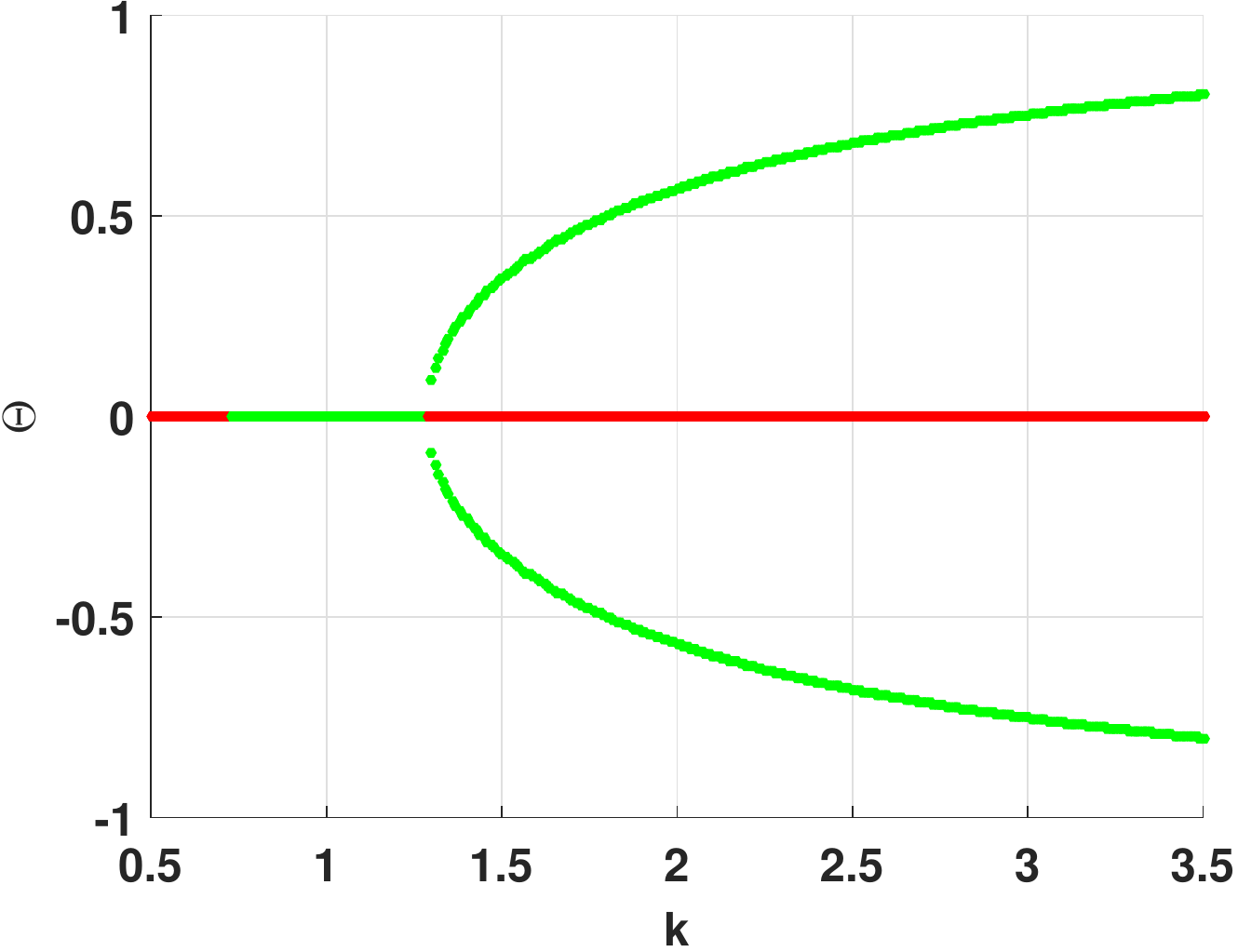}
	}
\caption{ The spontaneous symmetry breaking of patterns supported by the
double-peak spatial modulation of the nonlinearity. Asymmetry parameter (%
\protect\ref{Theta}) is displayed vs. propagation constant $k$, as obtained
from the numerical solution of Eq. (\protect\ref{u_model_1d_single_peak})
with $\protect\alpha =0.5$, $\protect\beta =0$, and $g(x)$ taken as per the
second line in Table \protect\ref{Table1} with $\Delta =0.244$. The three
panels correspond to different strengths of the background nonlinearity: $%
\protect\sigma =-1$ (a), $\protect\sigma =0$ (b), and $\protect\sigma =+1$
(c). Green and red branches represent stable and unstable solutions,
respectively. Red segments of the symmetric solutions ($\Theta =0$) at small
$k$ correspond to very weak(actually, formal) instability of modes with the
convex shape, see below.}
\label{fig1}
\end{figure*}

\begin{figure*}[tbp]
\centering\captionsetup[subfigure]{labelformat=empty}
\subfloat[]{
		\includegraphics[width=0.8\textwidth]{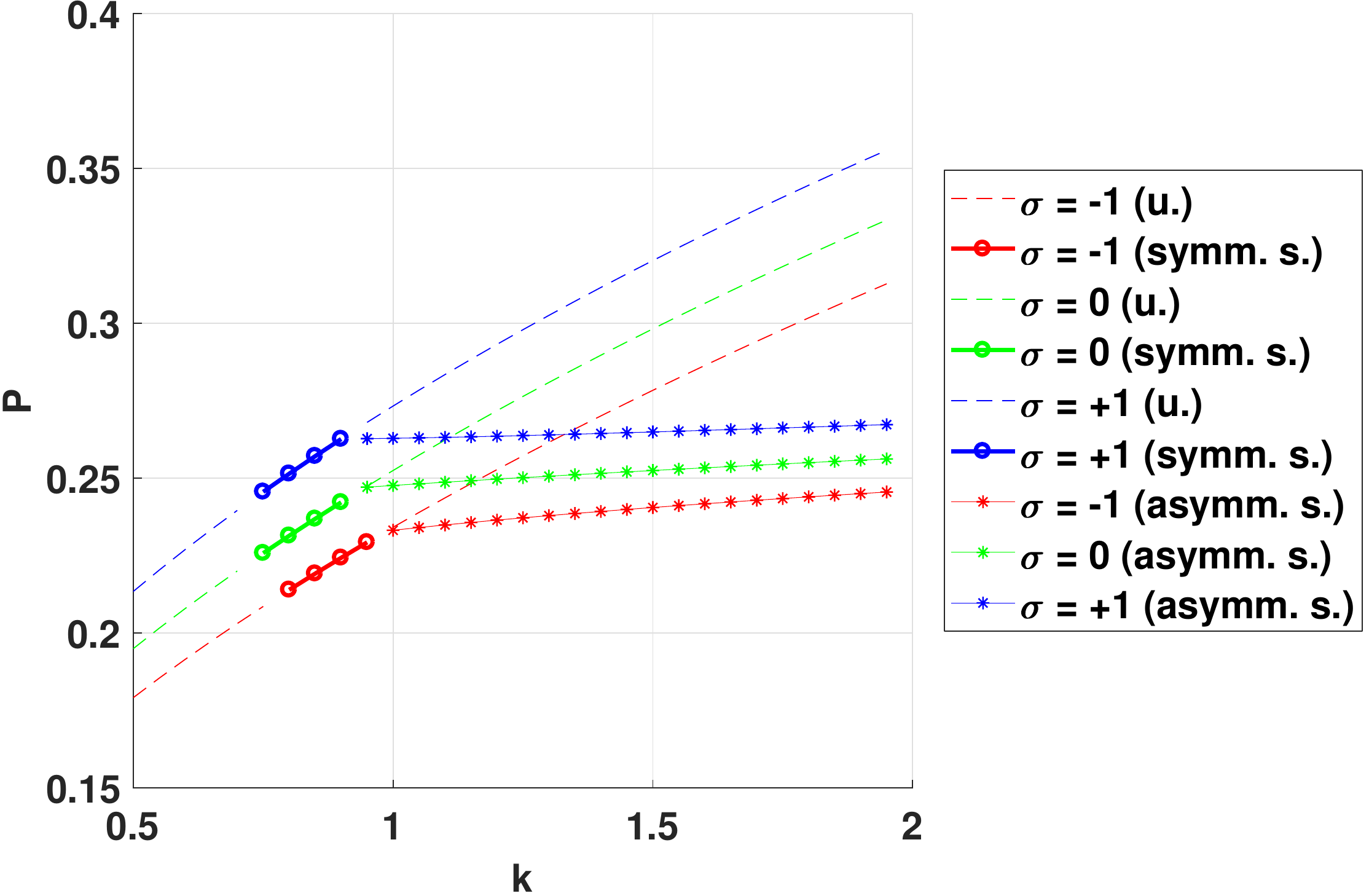}
	}
\caption{The total power as a function of the propagation constant, $k$, for
families of symmetric and asymmetric states in the system with the
double-peak singular modulation of the local nonlinearity, for $\Delta
=0.293 $ in the second line of Table \protect\ref{Table1}, $\protect\alpha %
=0.5$, $\protect\beta =0$ and different values of $\protect\sigma $ in Eq. (%
\protect\ref{u_model_1d_single_peak}). Here, and in Fig. \protect\ref{fig6}
below, symbols in the notation box mean: ``u." - unstable symmetric
solutions, ``symm. s." - stable symmetric ones, and ``asymm. s." - stable
asymmetric states. }
\label{fig2}
\end{figure*}

The symmetry breaking indeed happens with the increase of propagation
constant $k$, as shown in Figs. \ref{fig1} and \ref{fig2} for $\beta =0$ [no
linear potential in Eq. (\ref{u_model_1d_single_peak})]. These figures
clearly demonstrate that the SSB in the present system corresponds to the
supercritical pitchfork bifurcation \cite{Iooss}, which destabilizes the
symmetric states, replacing them by stable asymmetric ones. A qualitatively
similar result was obtained for the nonlinearity modulation format
represented by a pair of symmetric Gaussians (regularized $\delta $%
-functions) in \cite{Dong}, which confirms the generic character of this
SSB\ scenario in systems with the double-peak modulation of the local
self-focusing nonlinearity. This conclusion is further corroborated by the
fact that the presence of the uniform background nonlinearity of either sign
does not essentially effect the SSB scenario, although it naturally shifts
the total power at the bifurcation point to larger values in the case of the
defocusing background, as seen in Fig. \ref{fig2}.

\begin{figure*}[tbp]
\centering
\subfloat[]{
\includegraphics[width=0.32\textwidth]{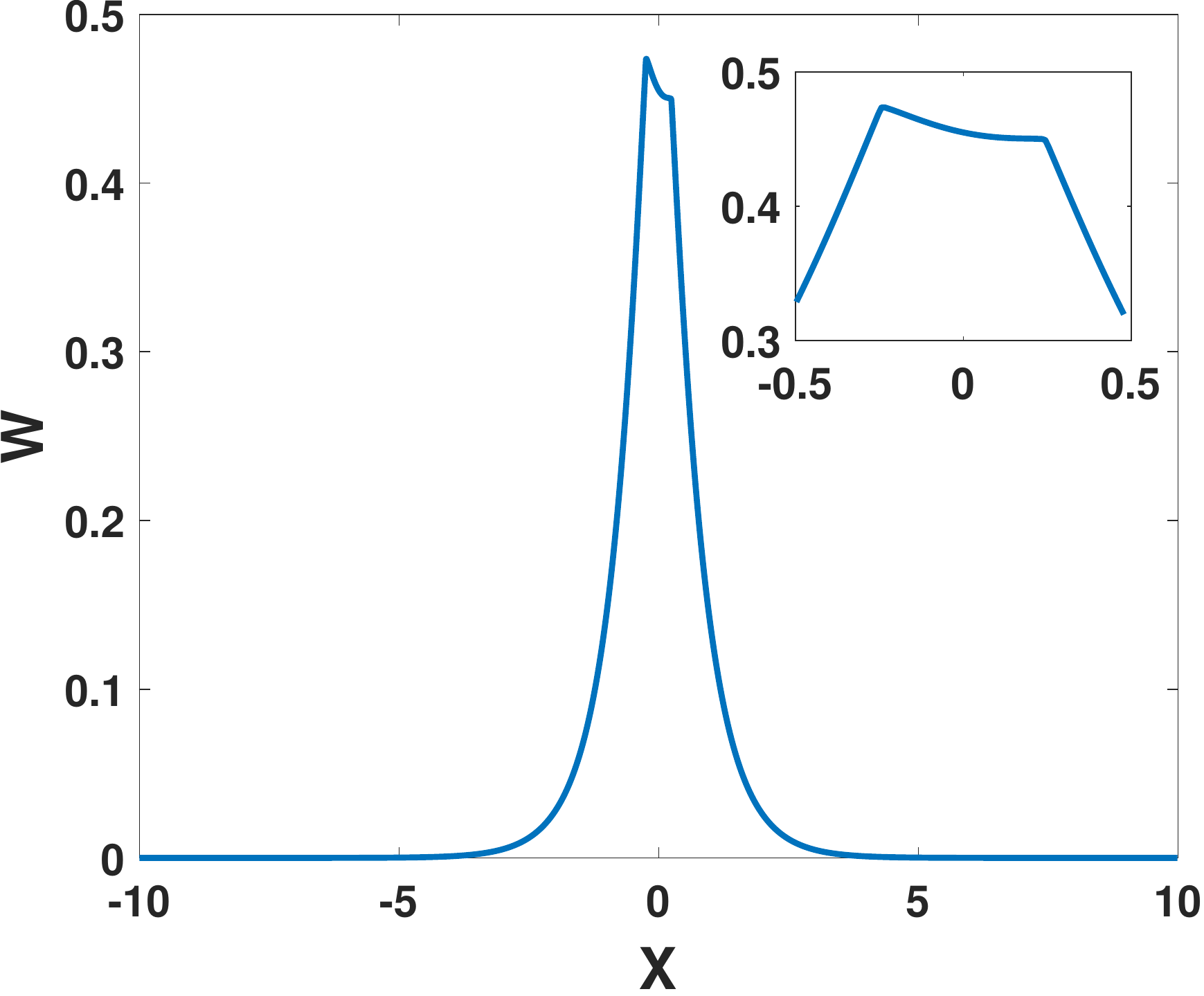}
} \subfloat[]{
\includegraphics[width=0.32\textwidth]{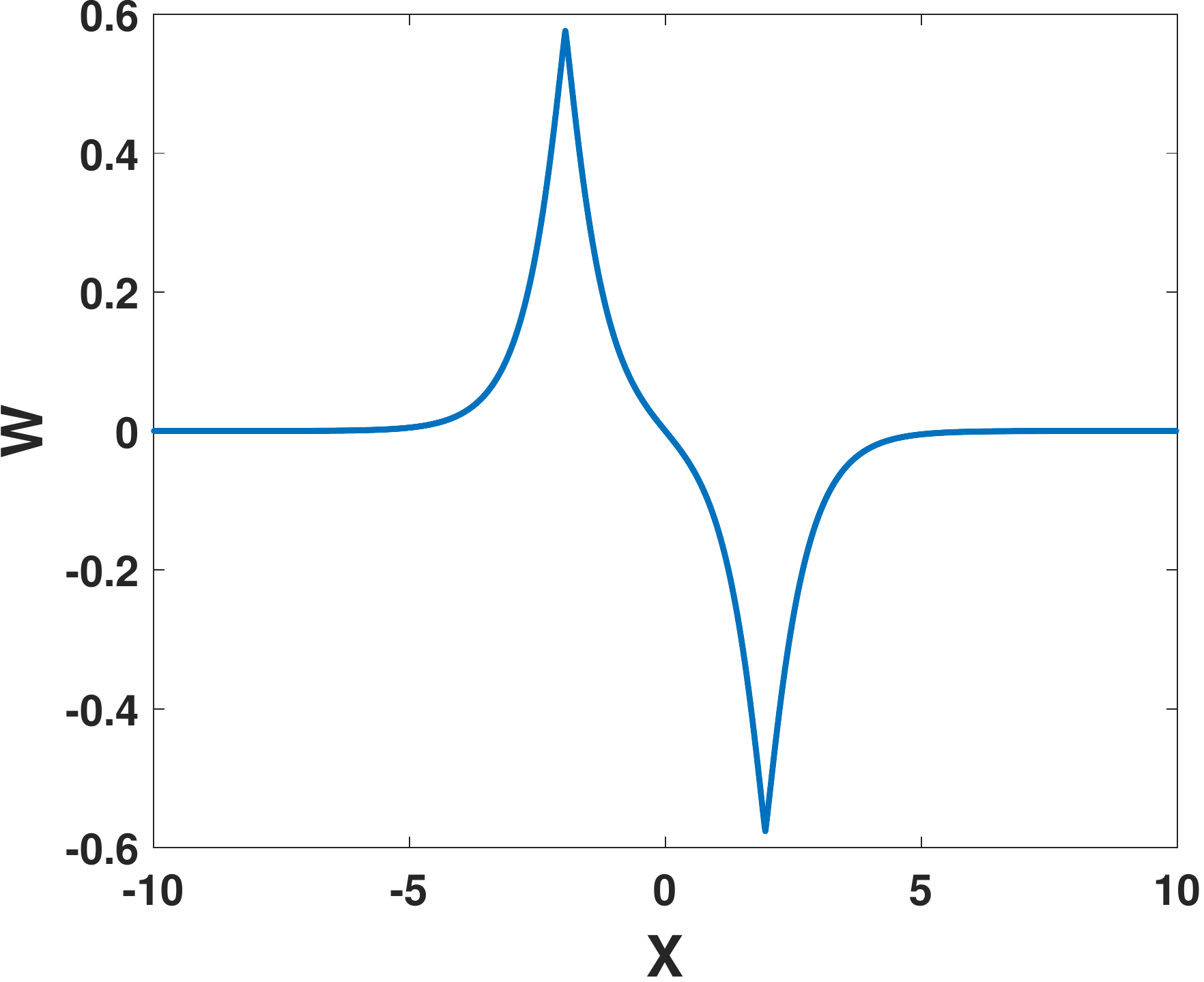}
} \subfloat[]{
\includegraphics[width=0.32\textwidth]{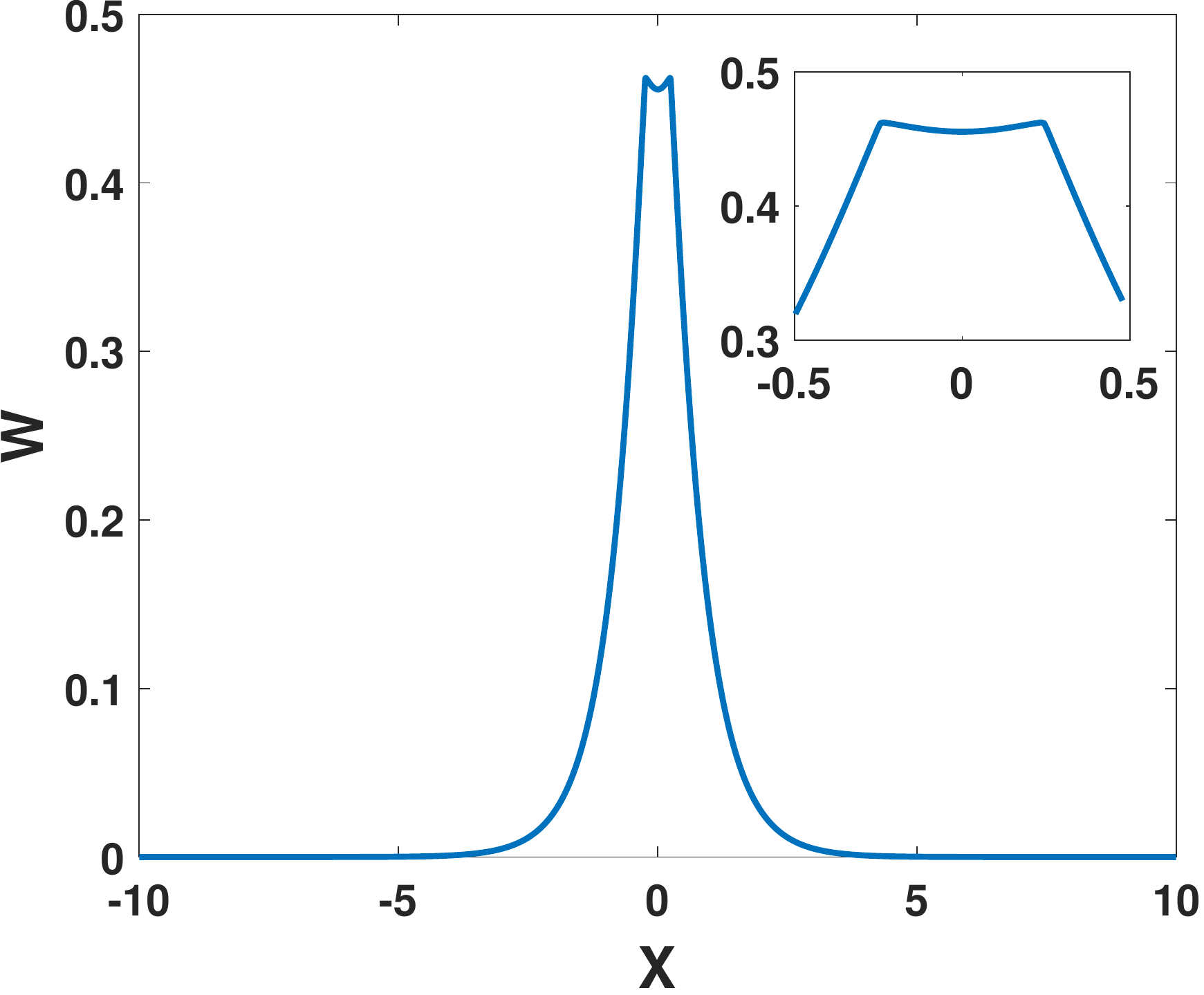}
} \newline
\par
\subfloat[]{
	\includegraphics[width=0.32\textwidth]{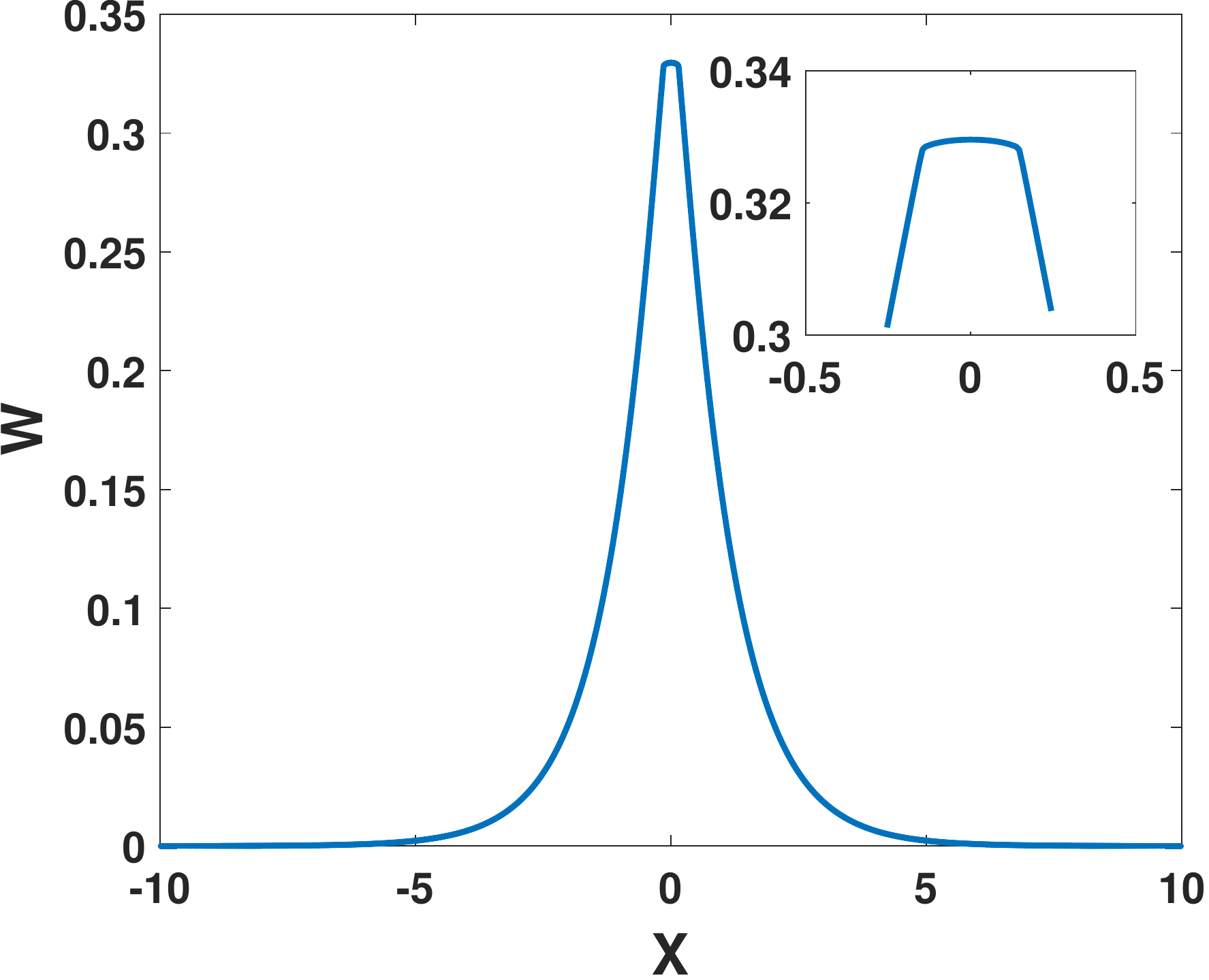}
} \subfloat[]{
	\includegraphics[width=0.32\textwidth]{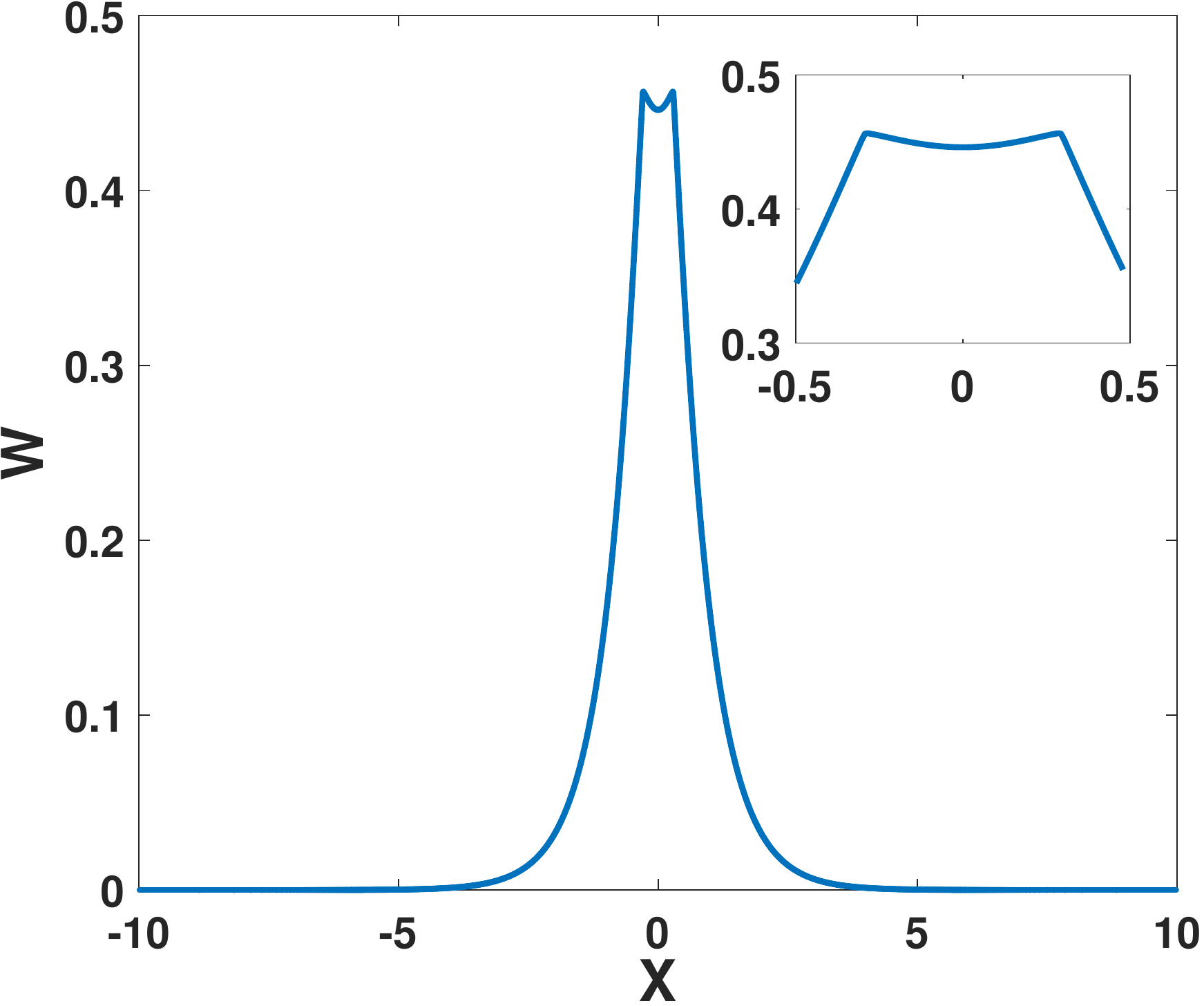}
}
\caption{Typical shapes of modes pinned to the double-peak
nonlinearity-modulation pattern. Panels (a), (b), and (c) display,
respectively, a stable weakly asymmetric state with $\Theta =0.0525$, an
unstable twisted (antisymmetric) one, and an unstable symmetric state, for $%
\Delta =0.244$, and a fixed value of the propagation constant, $k=1.4$. (d)
An example of a formally (very weakly) unstable symmetric state with the
convex shape, for $\Delta =0.1465$ and $k=0.55$. (e) A completely stable
symmetric state with the concave shape, for the $\Delta =0.293$ and $%
k=1.3$. Insets in (d) and (e) zoom the convex and concave top parts of the
respective profiles. In all the panels, other parameters in Eq. (\protect\ref%
{u_model_1d_single_peak}) are $\protect\alpha =0.5,$ $\protect\sigma =-1,$
and $\protect\beta =0$.}
\label{fig3}
\end{figure*}

\begin{figure*}[tbp]
\captionsetup[subfigure]{labelformat=empty} \centering%
\subfloat[]{
		\includegraphics[width=0.5\textwidth]{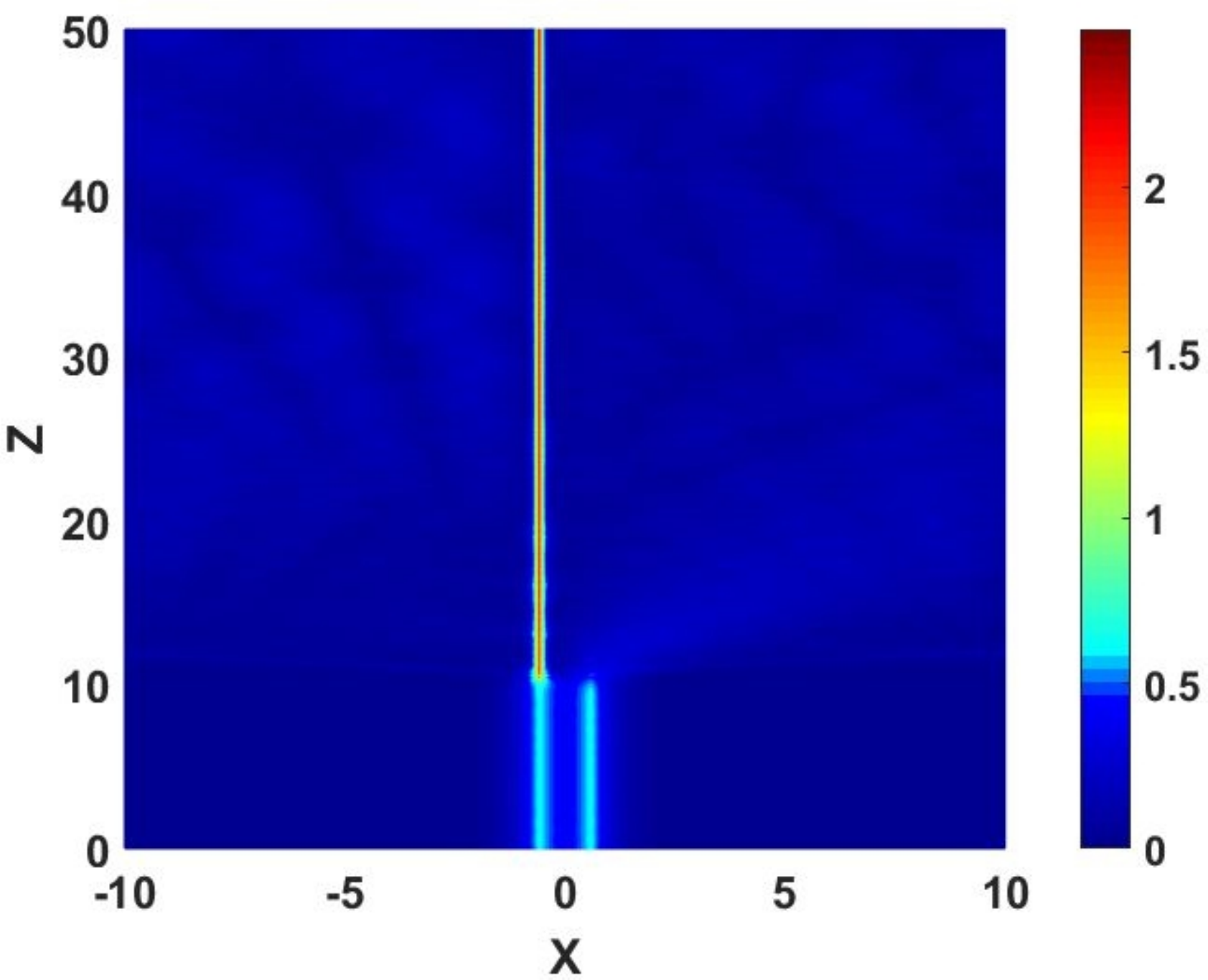}
	}
\caption{The spontaneous transformation of an unstable symmetric state into
a stable asymmetric one in the double-peak system, for $\protect\alpha =0.5$%
, $\protect\sigma =-1$, $\Delta =0.586$, and $k=2$. }
\label{fig4}
\end{figure*}

Generic examples of stationary symmetric and asymmetric modes are displayed
in Figs. \ref{fig3}(c) and \ref{fig3}(a), respectively. The conclusion about
the destabilization of the symmetric states at the SSB point, which was
produced by the calculation of stability eigenvalues by means of Eq. (\ref{G}%
), is confirmed by direct simulations, demonstrating that unstable symmetric
states spontaneously transform into their stable asymmetric counterparts, as
shown in Fig. \ref{fig4}. 


\begin{figure*}[tbp]
\centering
\subfloat[]{
		\includegraphics[width=0.49\textwidth]{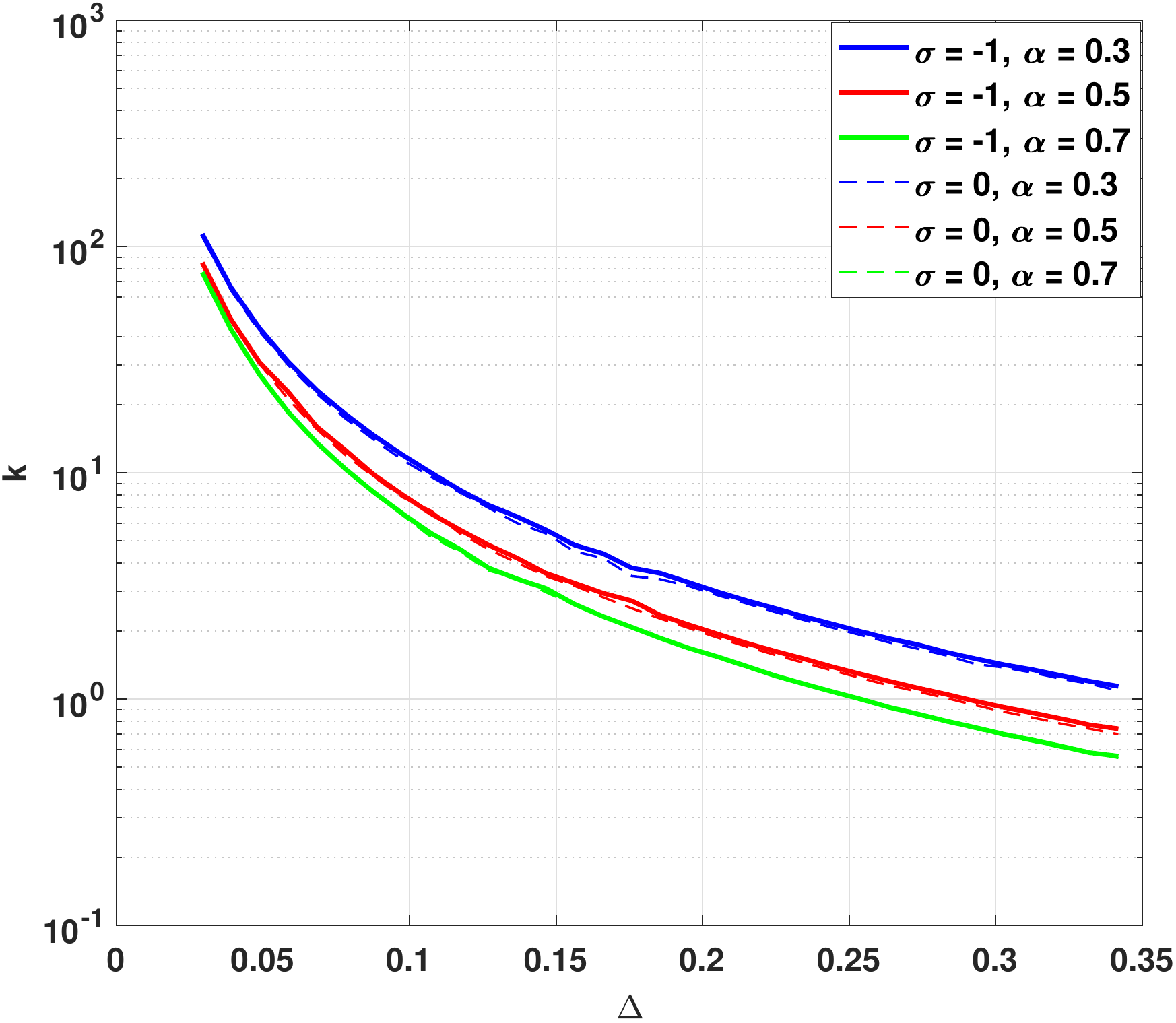}
	} \subfloat[]{
	\includegraphics[width=0.49\textwidth]{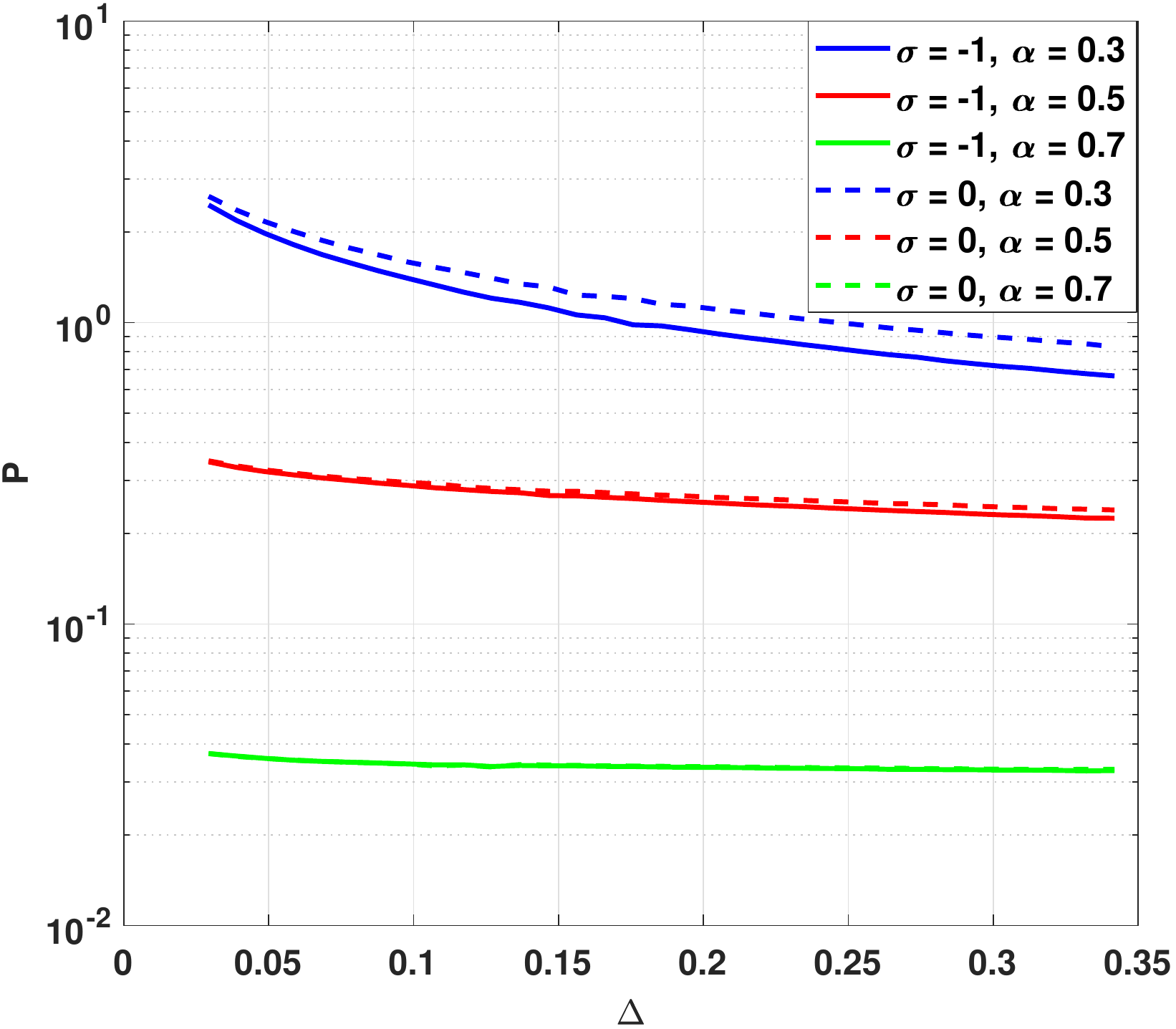}
	}
\caption{ Values of the propagation constant, $k$ (a), and total power, $P$
(b), at the point of the spontaneous-symmetry-breaking bifurcation, vs.
half-separation, $\Delta $, between the two nonlinearity-modulation peaks,
with $\protect\beta =0$. The dependences are shown on the log scale, for
different values of the singular-modulation power, $\protect\alpha $, and
background-nonlinearity strength, $\protect\sigma $.}
\label{fig5}
\end{figure*}

Data concerning the location of the SSB (i.e., bifurcation) points are
collected in Fig. \ref{fig5}, which provides a comprehensive map of the
parameter space of the double-peak setting. It is seen that, as mentioned
above, the dependence of the critical values of the propagation constant ($k$%
) and total power ($P$) on the strength of the background nonlinearity, $%
\sigma $, is quite weak, while the dependence on the modulation's
singularity degree, $\alpha $ [see Table \ref{Table1}], is essential.
Naturally, $k\rightarrow \infty $ and $P\rightarrow \infty $ is observed
Fig. \ref{fig5} at $\Delta \rightarrow 0$, as the SSB may happen if the
width of the mode pinned to an individual peak, which scales as $k^{-1/2}$,
has the same order of magnitude as $\Delta $.

\begin{figure*}[tbp]
\captionsetup[subfigure]{labelformat=empty} \centering
\subfloat[]{
		\includegraphics[width=0.8\textwidth]{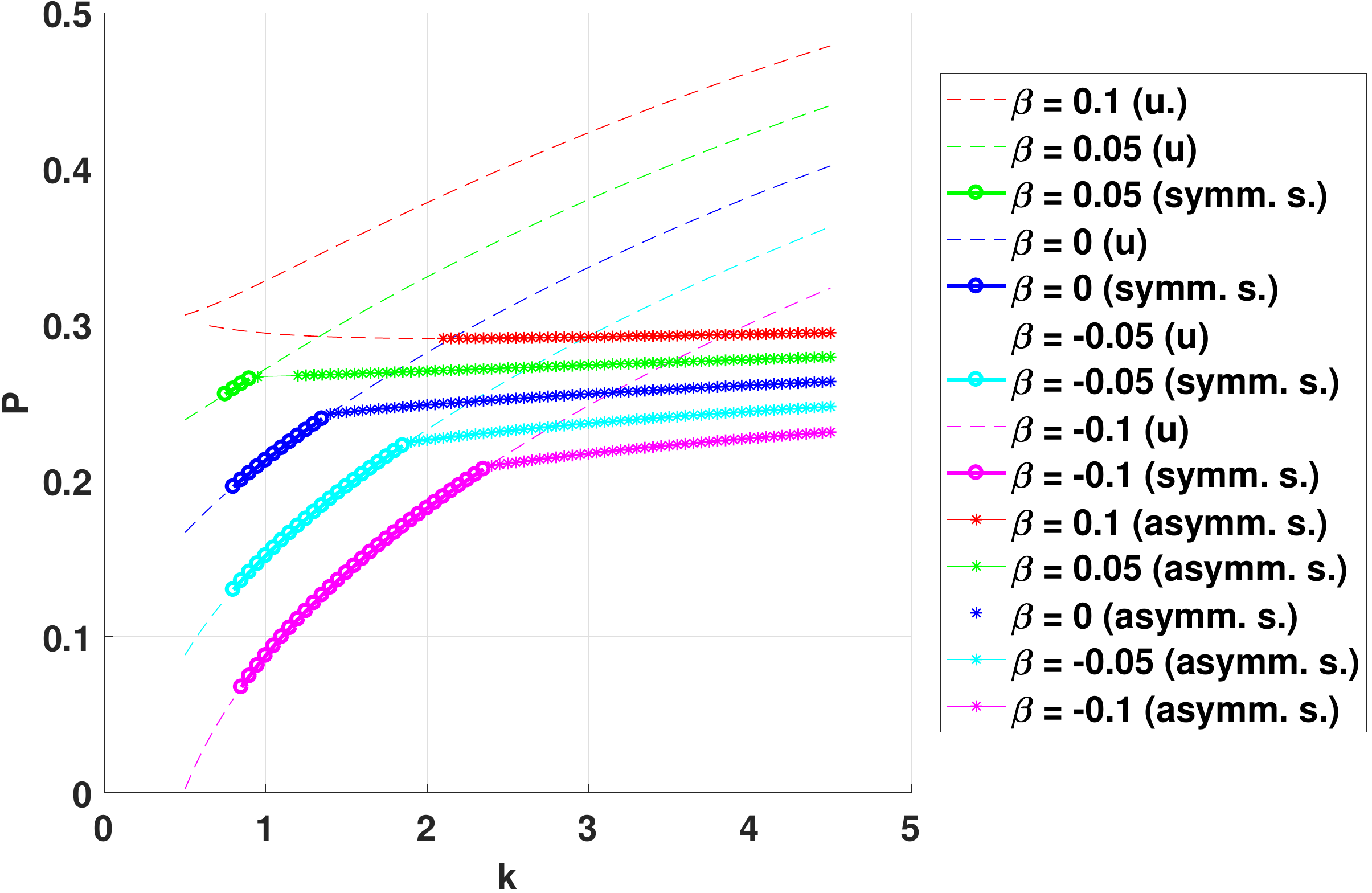}
	}
\caption{The same as in Fig. \protect\ref{fig2}, but for $\protect\alpha %
=0.5 $, $\protect\sigma =-1$, $\Delta =0.244$, and different values of $%
\protect\beta $. It is seen that, in the presence of $\protect\beta >0$,
weakly asymmetric stationary states become partially unstable (see, e.g.,
the solution families for $\protect\beta =0.05$, where a gap between chains
of green circles and stars corresponds to the instability). The instability
leads to transformation of the weakly asymmetric modes into 
asymmetric breathers, as shown in Fig. \protect\ref{fig7}(a)}.
\label{fig6}
\end{figure*}

\begin{figure*}[tbp]
\centering
\subfloat[]{
		\includegraphics[width=0.49\textwidth]{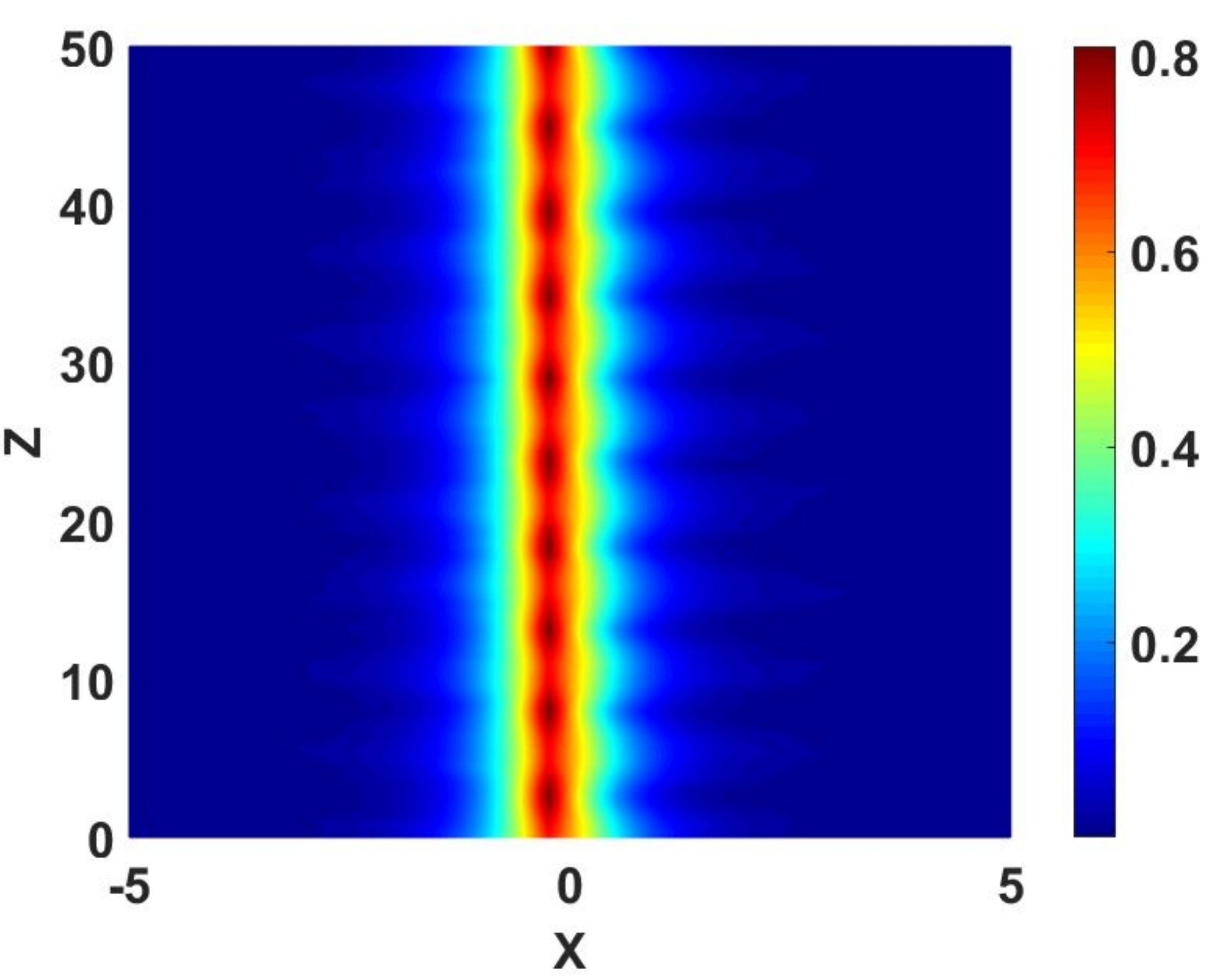}}
	\subfloat[]{
		\includegraphics[width=0.49\textwidth]{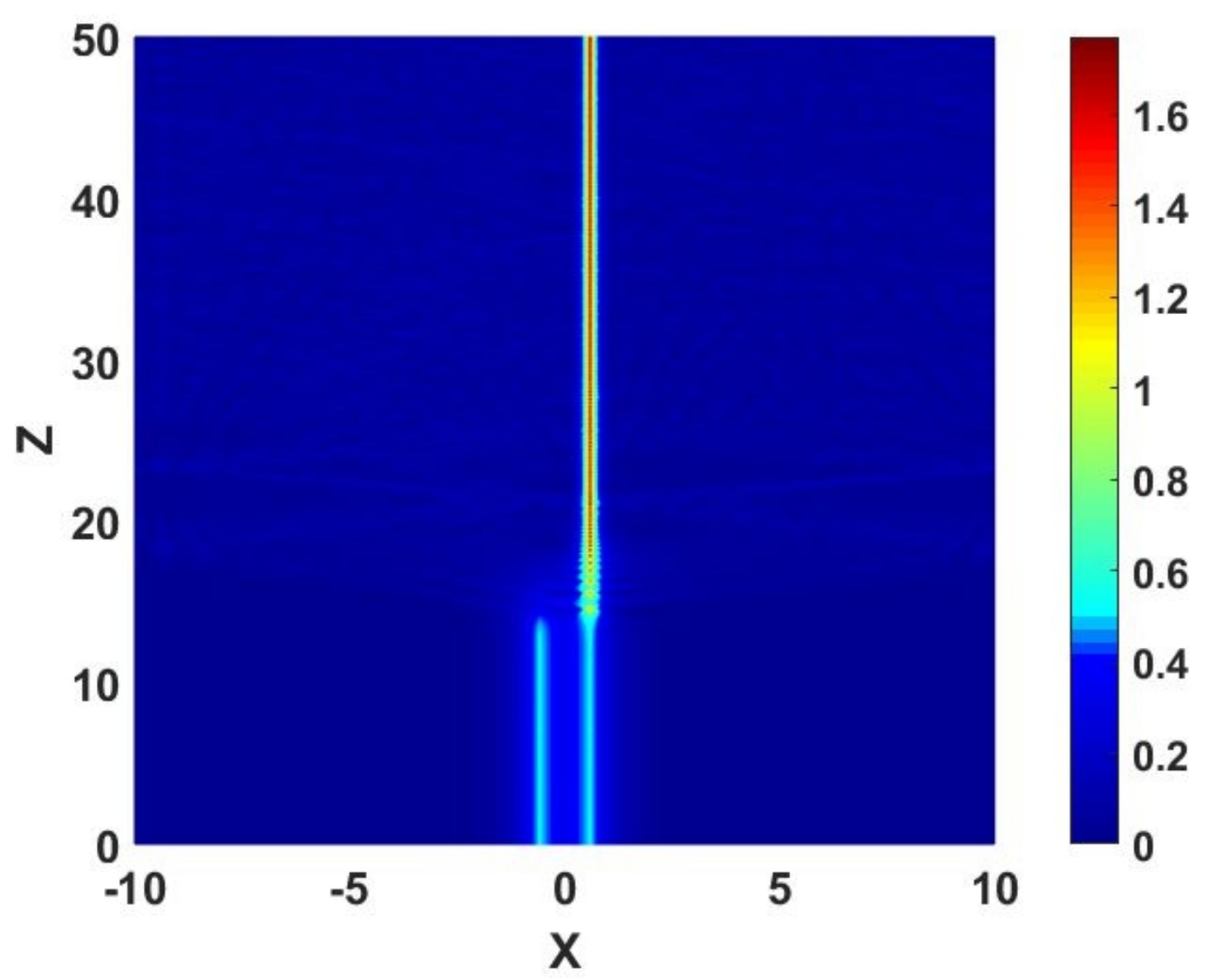}
	}
\caption{The development of the instability of (a) unstable asymmetric state
with $\protect\Theta = 0.4043$ for $\protect\alpha =0.5$, $\Delta =0.2441$, $\protect\sigma =-1$, and $k=1.2$ in the presence of the repulsive linear-potential component in the singular-modulation double peak, corresponding to $\protect\beta =0.1$
and (b) an unstable symmetric state, for $\protect\alpha =0.5$, $\Delta =0.586$, $\protect\sigma =-1$, and $k=2$, in the presence of the repulsive linear-potential component corresponding to $\protect\beta =-0.1$.}
\label{fig7}
\end{figure*}

Unlike the uniform background nonlinearity, the presence of the linear
component of the modulation peaks, represented by coefficient $\beta $ in
Eq. (\ref{eq:u_model_1d}), produces a strong effect on the SSB scenario.
Namely, $\beta >0$ (which correspond to the repulsive linear potential)
leads to partial destabilization of the asymmetric states, as shown in Fig. %
\ref{fig6}. In this case, the pair of the singular-modulation peaks may support 
two different asymmetric states: one with a relatively small difference 
between amplitudes of the solution at the two peaks, and a fully asymmetric
state, strongly pinned to one of the peaks. The repulsive linear potential destabilizes the
weakly asymmetric state, converting it into an asymmetric breather, as shown in Fig. \ref{fig7}(a).
On the other hand, the unstable symmetric states are transformed into strongly 
asymmetric modes pinned to one of the peaks, as shown in Fig. \ref{fig7}(b).
This finding may be explained by the fact that the increase of the amplitude
of the pinned soliton helps the nonlinear attractive potential to overcome
the repulsion induced by the linear potential. On the other hand,
symmetric states, unlike the asymmetric ones, are not destabilized by the linear component of the modulation peaks.

\begin{figure*}[tbp]
\captionsetup[subfigure]{labelformat=empty} \centering
\subfloat[]{
	\includegraphics[width=0.8\textwidth]{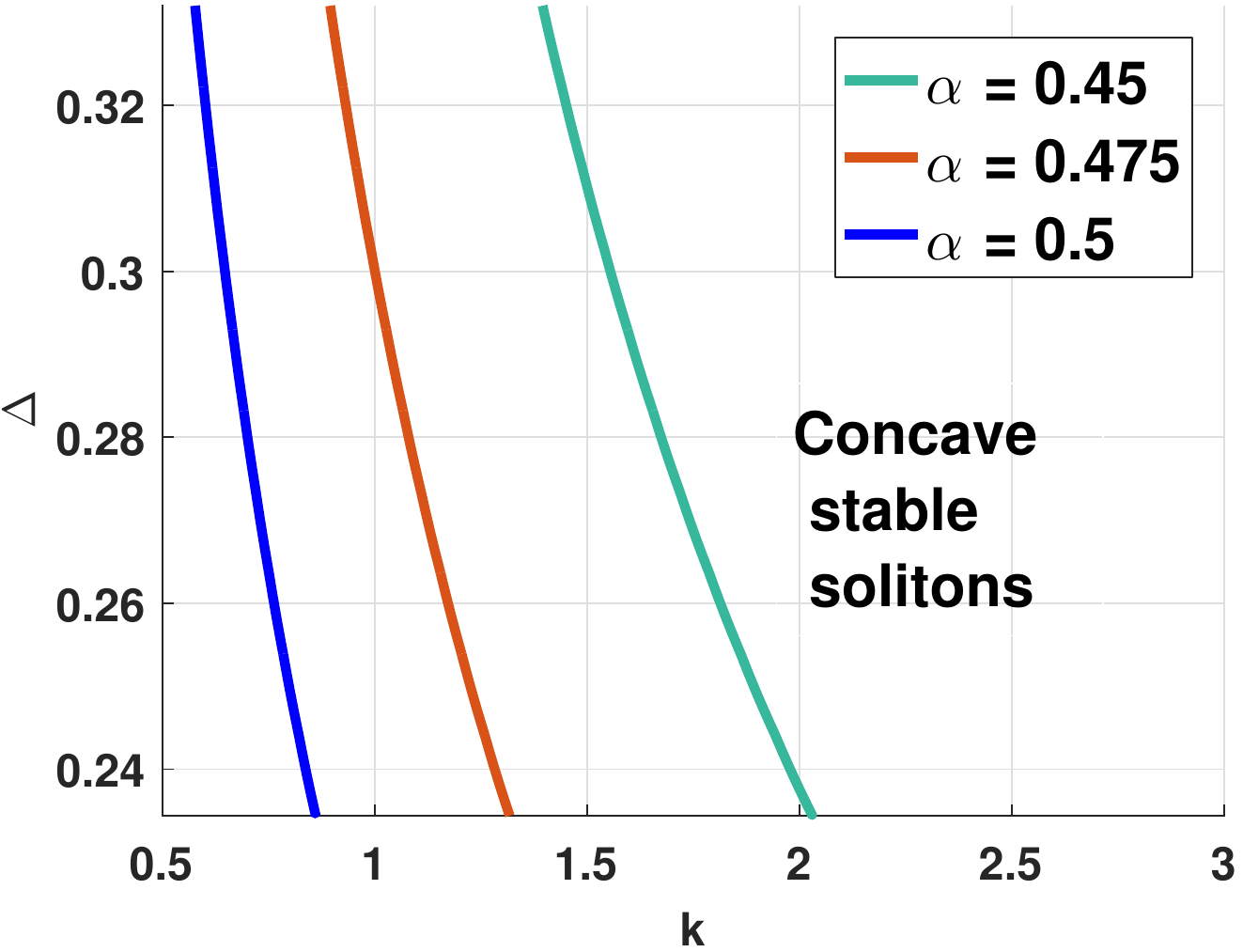}
}
\caption{Boundaries between areas occupied by strongly overlapped modes with
concave (completely stable) shapes, and concave ones, subject to a very weak
instability. Here, $\protect\sigma =-1$ and $\protect\beta =0$ are fixed,
the boundaries being shown for three different values of $\protect\alpha $,
as indicated in the figure. }
\label{fig8}
\end{figure*}

In the limit case when the localized modes pinned to adjacent nonlinearity
peaks are strongly overlapped, the two modes merge into a nearly flat-top
one, with the small deviation of the flatness corresponding to slightly
convex or concave shapes, as shown in Figs. \ref{fig3}(d) and \ref{fig3}(e),
respectively. Further analysis demonstrates that the concave shapes are
completely stable, while the computation of the eigenvalues of operator (\ref%
{G}) for convex ones (which appear when the two individual modes become
completely overlapping) features a very weak instability, that corresponds
to short red segments near the left edge of all the panels in Fig. \ref{fig1}%
. The eigenvalues that characterize the instability are limited to be $%
\lesssim 10^{-4}$. Actually, direct simulations do not reveal any tangible
instability of the convex profiles, in accordance with the fact that the
corresponding eigenvalues are extremely small. Boundaries between the nearly
flat-top convex and concave profiles in the plane of $\left( k,\Delta
\right) $ for different values of $\alpha $ are displayed in Fig. \ref{fig8}%
. It is also relevant to mention that, in accordance with the results of
\cite{borovkova2012solitons}, where it was demonstrated that modes pinned to
a single modulation peak, competing with the self-defocusing uniform
background ($\sigma =+1$), exist if their total power exceeds a certain
threshold (minimum) value, the same effect was found here for the symmetric
modes, the threshold being virtually exactly equal to twice the one found in
\cite{borovkova2012solitons} (this feature is not shown in detail here).

\begin{figure*}[tbp]
\centering
\subfloat[]{
		\includegraphics[width=0.49\textwidth]{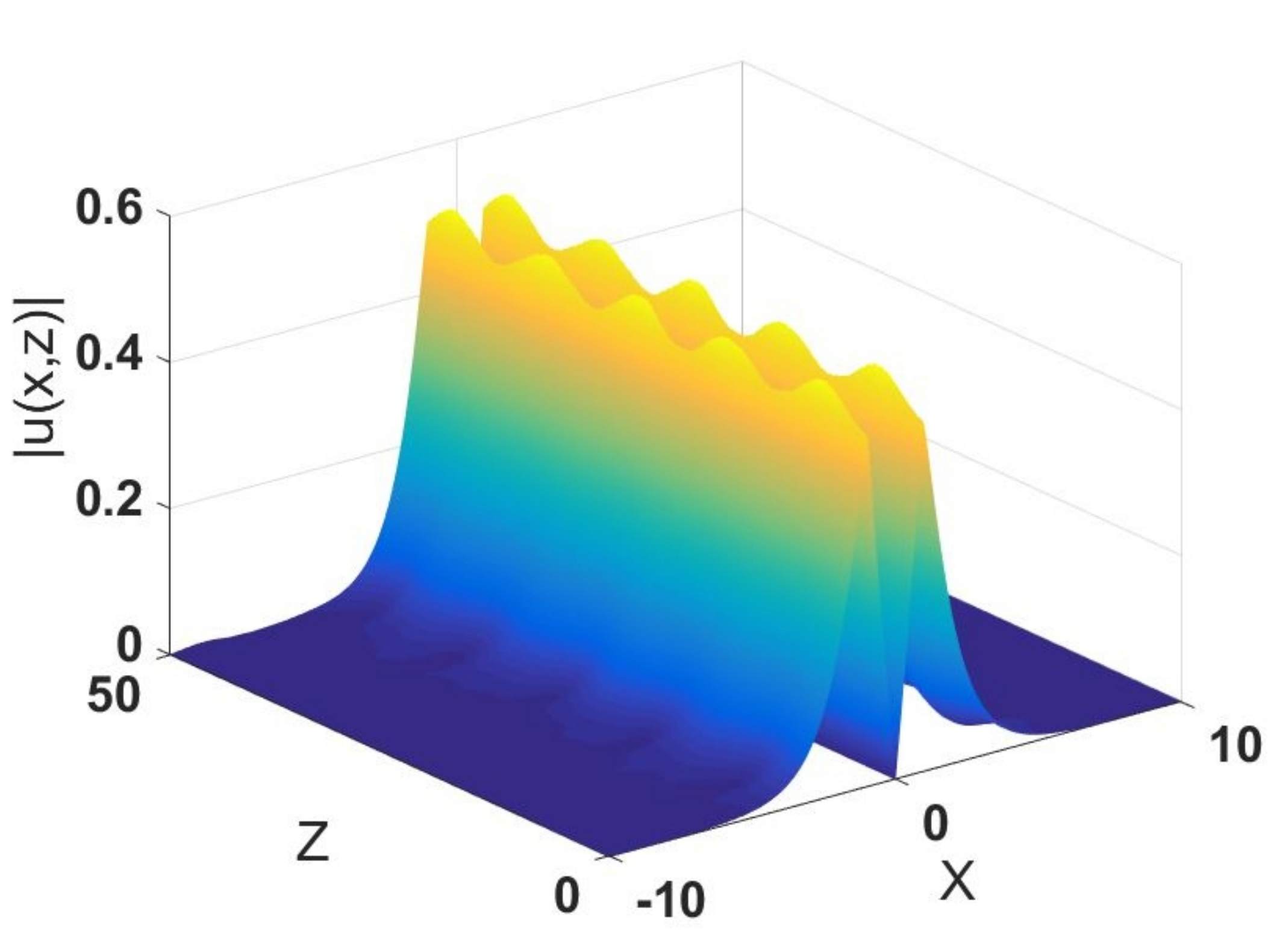}
	} \subfloat[]{
		\includegraphics[width=0.49\textwidth]{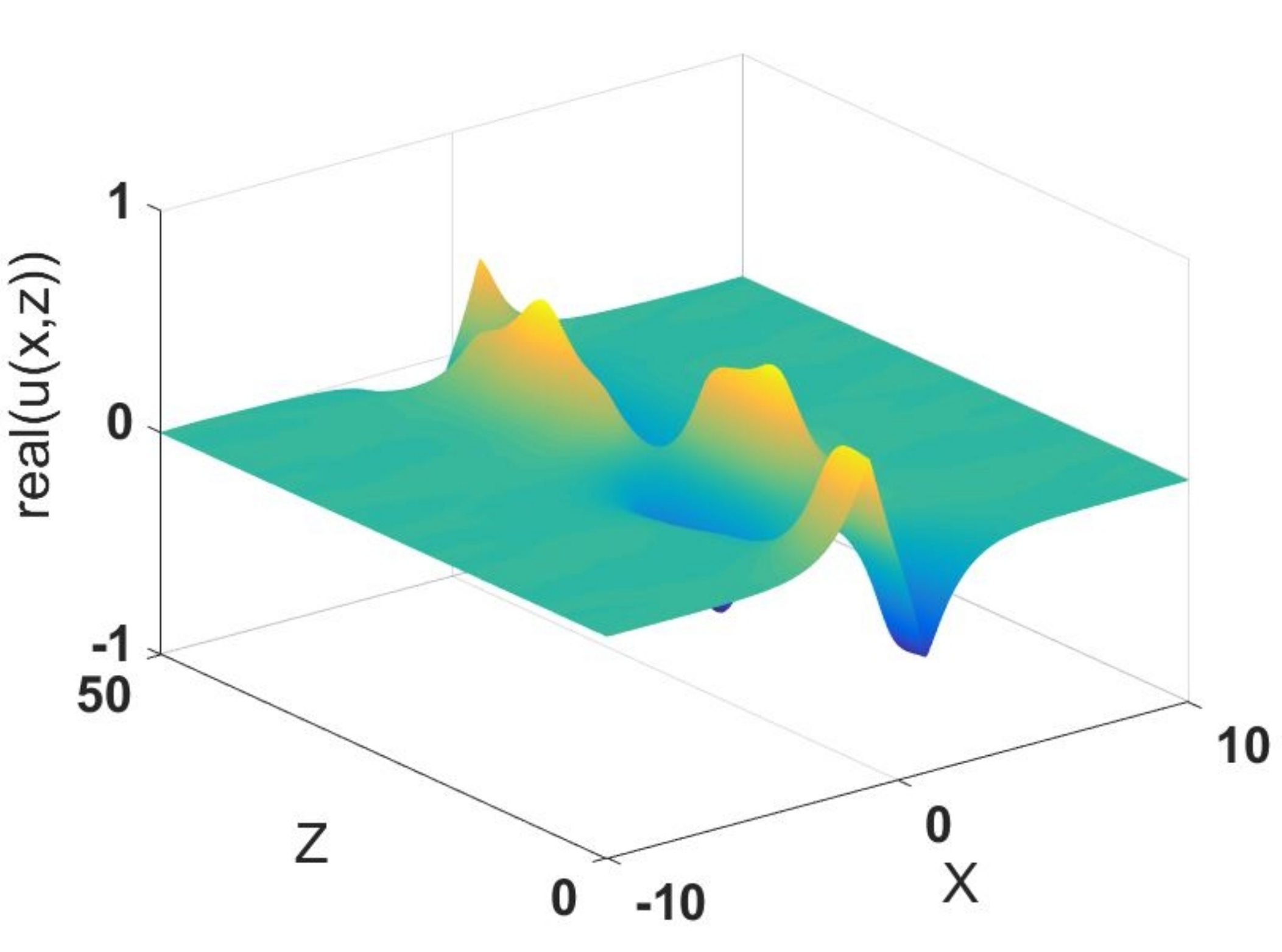}
	}
\caption{The evolution of an antisymmetric (twisted) mode, for $\protect%
\alpha =0.5$, $\protect\sigma =-1$, $\Delta =0.9766$, $\protect\beta =0$,
and $k=0.55$. (a) The plot of $\left\vert u\left( x,z\right) \right\vert $
displays the spontaneous transformation of the unstable twisted model into a
breather. (b) The plot of the real part of $u(x,z)$ demonstrates that the
breather keeps the original antisymmetric structure.}
\label{fig9}
\end{figure*}

Lastly, the model with two singular-modulation peaks admits solutions for
antisymmetric (twisted) pinned modes too. However, both the computation of
the stability eigenvalues and direct simulations reveal that all the twisted
states are unstable. As shown in Fig. \ref{fig9}, the instability transforms
them into robust breathers, which preserve the twisted structure.

\subsection{\textbf{The system with three and five singular-modulation peaks
}}

\begin{figure*}[tbp]
	\centering
	\subfloat[]{
		\includegraphics[width=0.49\textwidth]{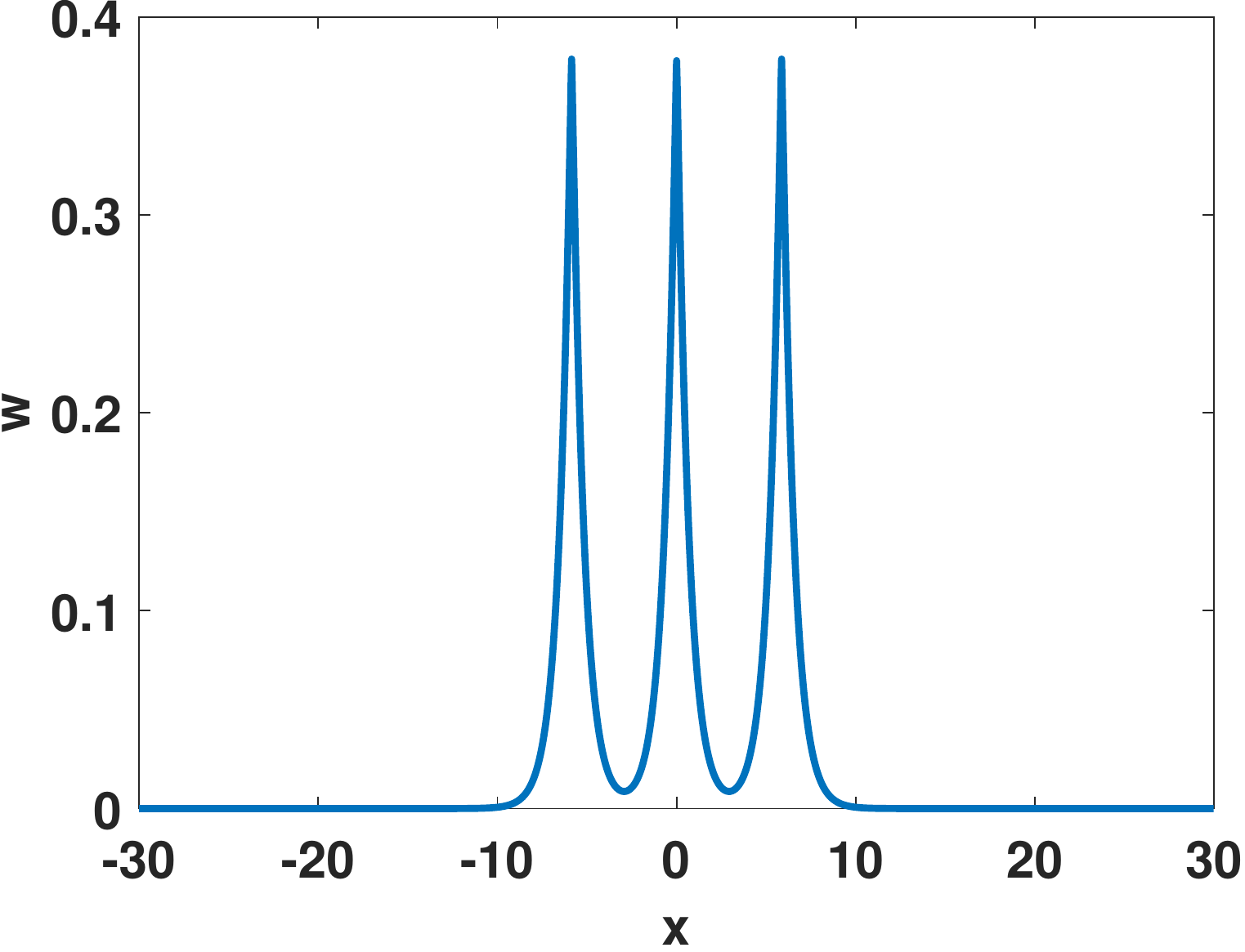}
	} \subfloat[]{
		\includegraphics[width=0.49\textwidth]{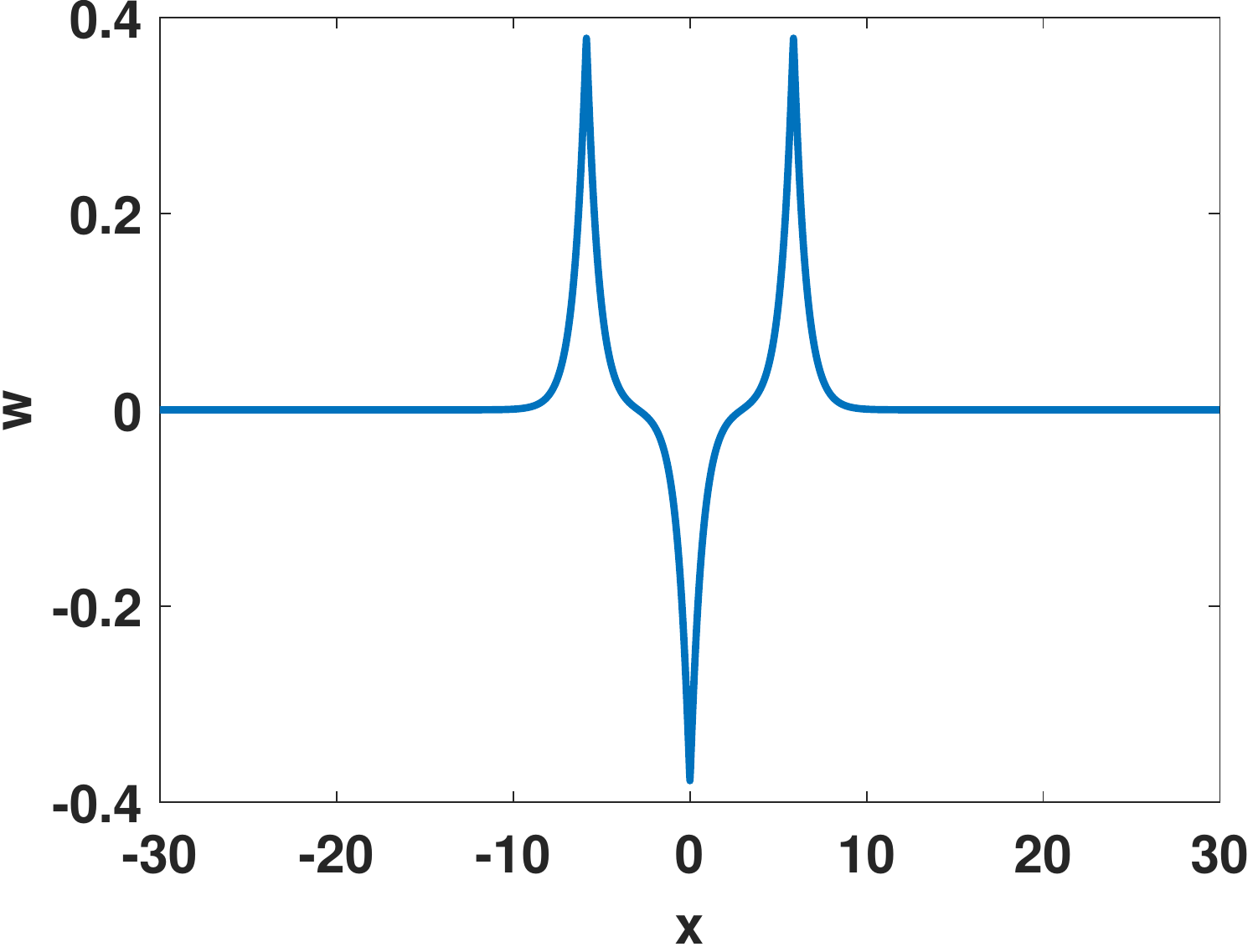}
	}
	\caption{ Modes pinned to the set of three singular-modulation peaks, which
		corresponds to the third line in Table \protect\ref{Table1}, with $\Delta
		=5.86$, and $\protect\sigma =-1$, $\protect\beta =0$ in Eq. (\protect\ref%
		{eq:u_model_1d}). (a) An unstable in-phase mode, with well-separated peaks
		[cf. the overlapped in-phase mode in Fig. \protect\ref{fig11}(b)]. (b) A
		stable twisted mode. Both modes were obtained for propagation constant $%
		k=1.2 $.}
	\label{fig10}
\end{figure*}

\begin{figure*}[tbp]
	\centering%
	\subfloat[]{
		\includegraphics[width=0.49\textwidth]{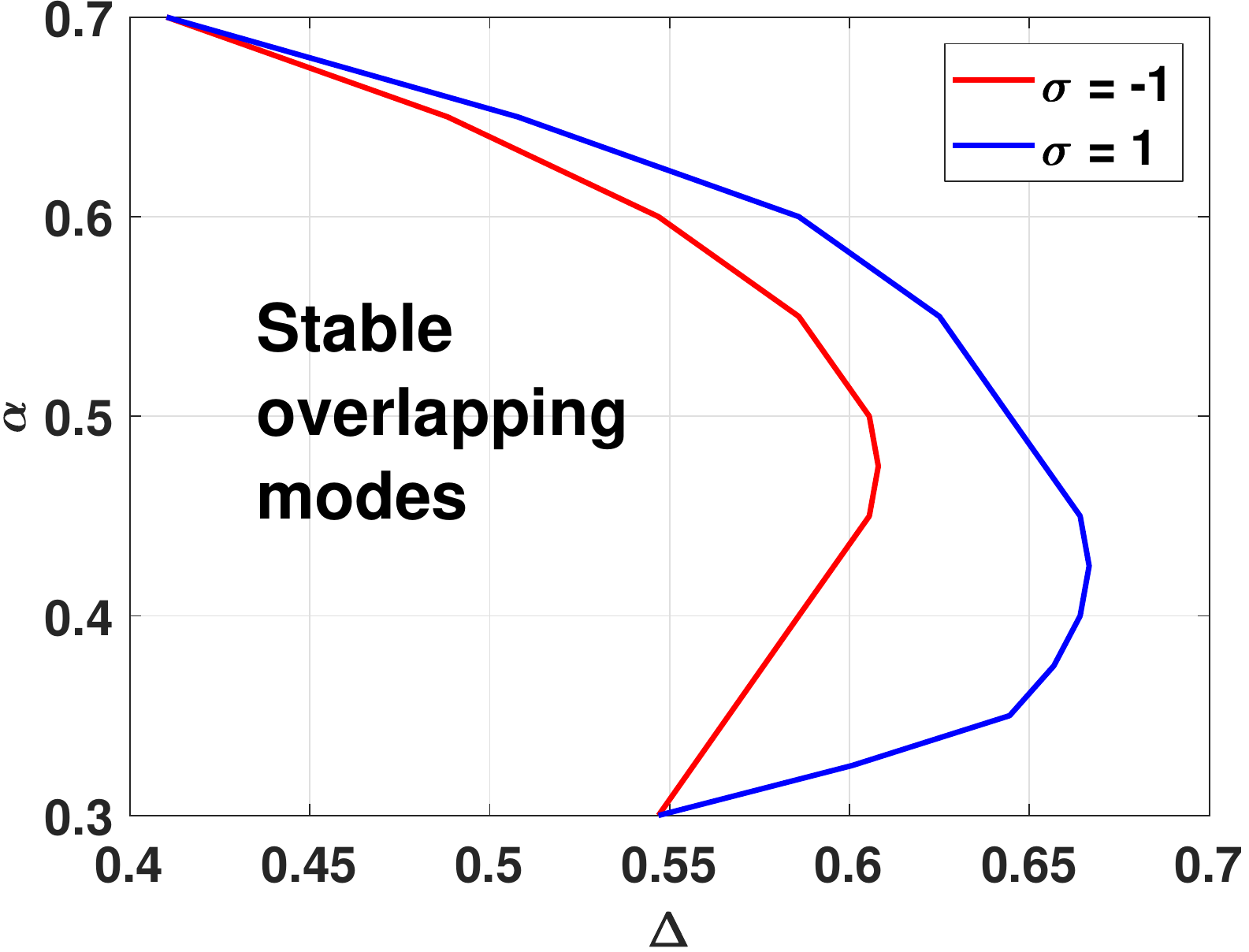}
	}%
	\subfloat[]{
		\includegraphics[width=0.49\textwidth]{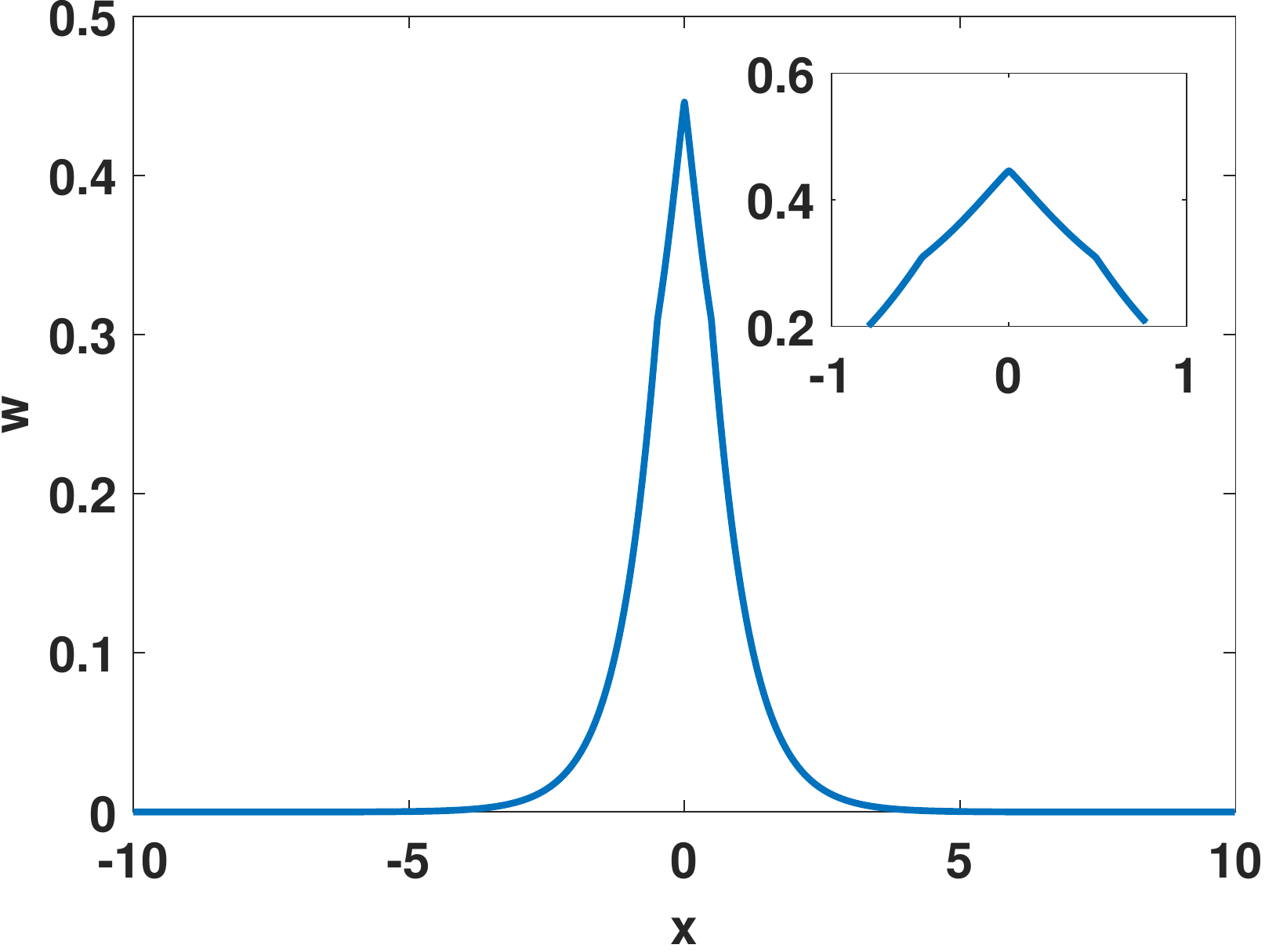}
	}
	\caption{ (a) Stability regions for strongly overlapping in-phase symmetric
		modes pinned by the triple singular-modulation peak, in the plane of $\left(
		\Delta ,\protect\alpha \right) $ for $\protect\sigma =\pm 1$, while other
		parameters are $\protect\beta =0$ and $k=1.2$. (b) A typical example of the
		stable mode, found at $\Delta =0.4883$ and $\protect\alpha =0.5$. The inset
		zooms the shape of the mode in the region where it covers the three
		singularity peaks.}
	\label{fig11}
\end{figure*}

The transition from the double singular-modulation peaks to a triple one
(which corresponds to the third line in Table \ref{Table1}) is of obvious
interest, as it makes it possible to understand respective changes in the
variety of pinned modes with different symmetries. In this case, asymmetric
modes (stable or unstable ones) have not been found, while symmetric states
exist in two varieties, \textit{viz}., with in-phase and out-of-phase (alias
twisted) shapes, which are shown in Fig. \ref{fig10}. The computation of the
corresponding eigenvalues, as well as direct simulations, demonstrates that
the in-phase modes with $k=1.2$ have their stability regions at $\Delta
<0.66 $ and $\Delta <0.61$ for $\sigma =+1$ and $-1$, respectively, as shown
in Fig. \ref{fig11}(a) (the stability\ regions have similar shapes at other
values of $k$). In this region, the modes are strongly overlapped, seeming
as a single local-power peak, see Fig. \ref{fig11}(b).

\begin{figure*}[tbp]
\captionsetup[subfigure]{labelformat=empty} \centering
\subfloat[]{
		\includegraphics[width=0.6\textwidth]{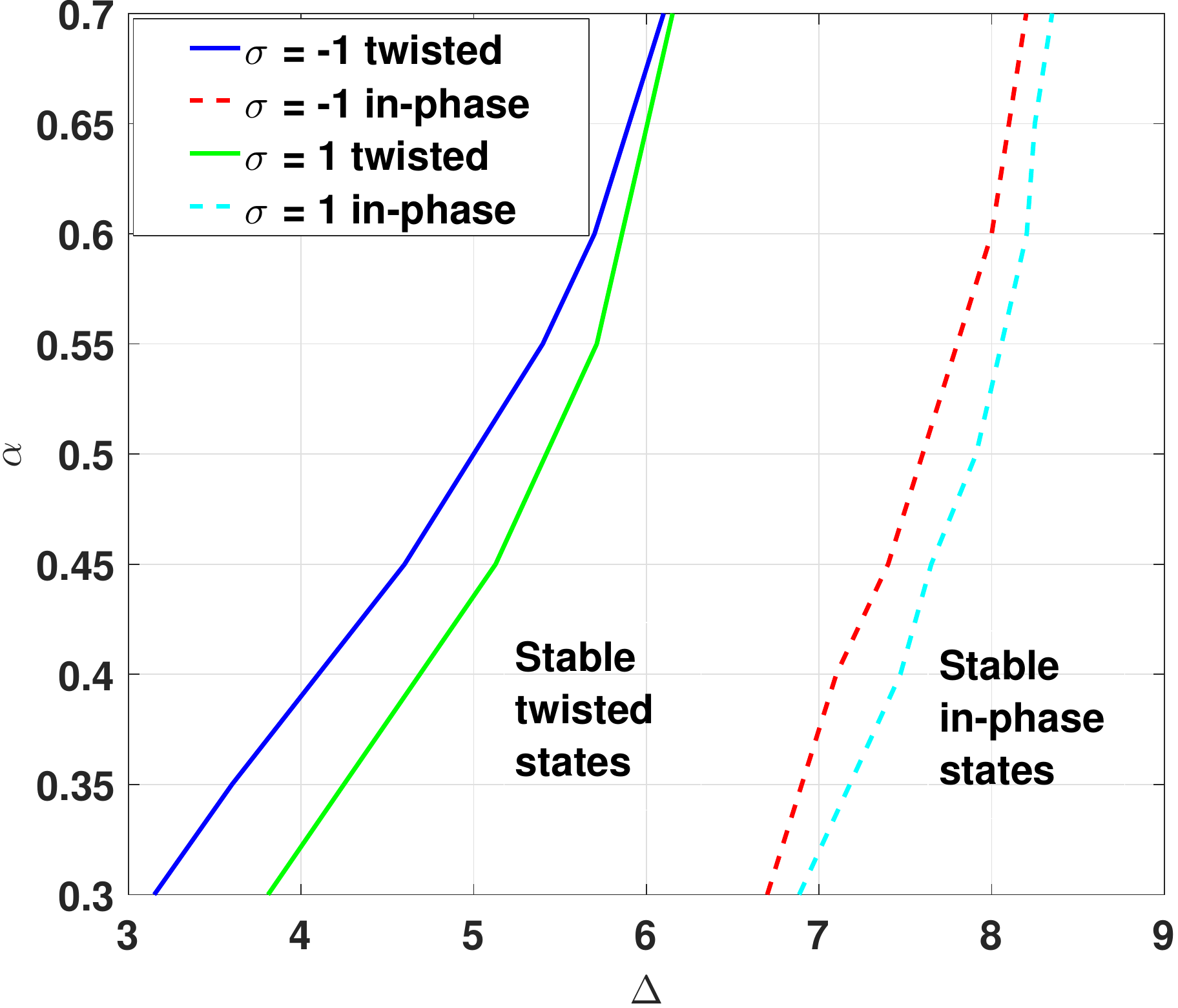}
	}
\caption{ Stability boundaries for twisted modes pinned to the set of three
peaks in the plane of $\left( \Delta ,\protect\alpha \right) $, represented
by the third line in Table \protect\ref{Table1} (the twisted modes are
unstable on the left-hand side of the continuous boundary lines). In this
figure, $\protect\beta =0$ and $k=1.2$ are fixed, while the strength of the
background uniform nonlinearity $\protect\sigma $ takes both values $\protect%
\sigma =+1$ and $-1$ (the self-defocusing and focusing background,
respectively). Beyond the dashed boundaries, in-phase modes cease being
unstable because the interaction between different modes pinned to
individual peaks becomes negligible.}
\label{fig12}
\end{figure*}

Outside of the above-mentioned regions, the in-phase states are always
unstable (unless separation $\Delta $ between the peaks is so large in
comparison to the width of local modes pinned to individual peaks that they
practically do not interact), while twisted modes have a nontrivial
stability area in the parameter plane of $\left( \Delta ,\alpha \right) $,
as shown in Fig. \ref{fig12}. It is worthy to note a conspicuous dependence
of the stability boundaries on strength $\sigma $ of the uniform
nonlinearity background in Figs. \ref{fig11} and \ref{fig12}, unlike a much
weaker dependence on $\sigma $ in the case of the double peak, cf. Fig. \ref%
{fig5}.

In direct simulations, all unstable modes pinned to the triple peak
spontaneously transform into patterns featuring either a state pinned to the
central peak, while the edge modes disappear, or two narrow modes pinned to
the edge peaks, while the central one disappears, see examples in Fig. \ref%
{fig13}. In the latter case, the  emerging pattern seems to be stable because
the narrow modes of which it is built virtually do not interact with each other.

\begin{figure*}[tbp]
\centering
\subfloat[]{
		\includegraphics[width=0.49\textwidth]{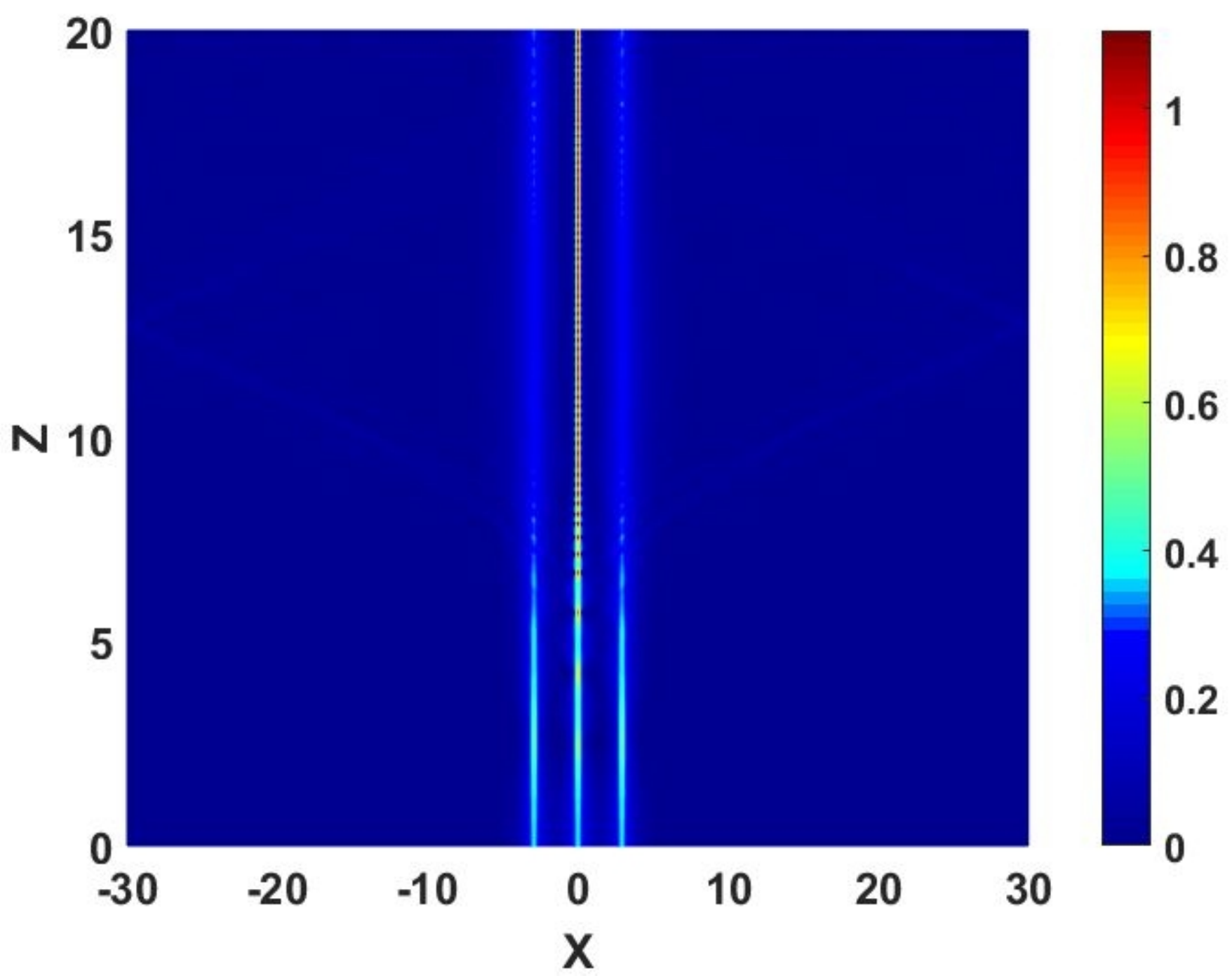}}
\subfloat[]{
	\includegraphics[width=0.49\textwidth]{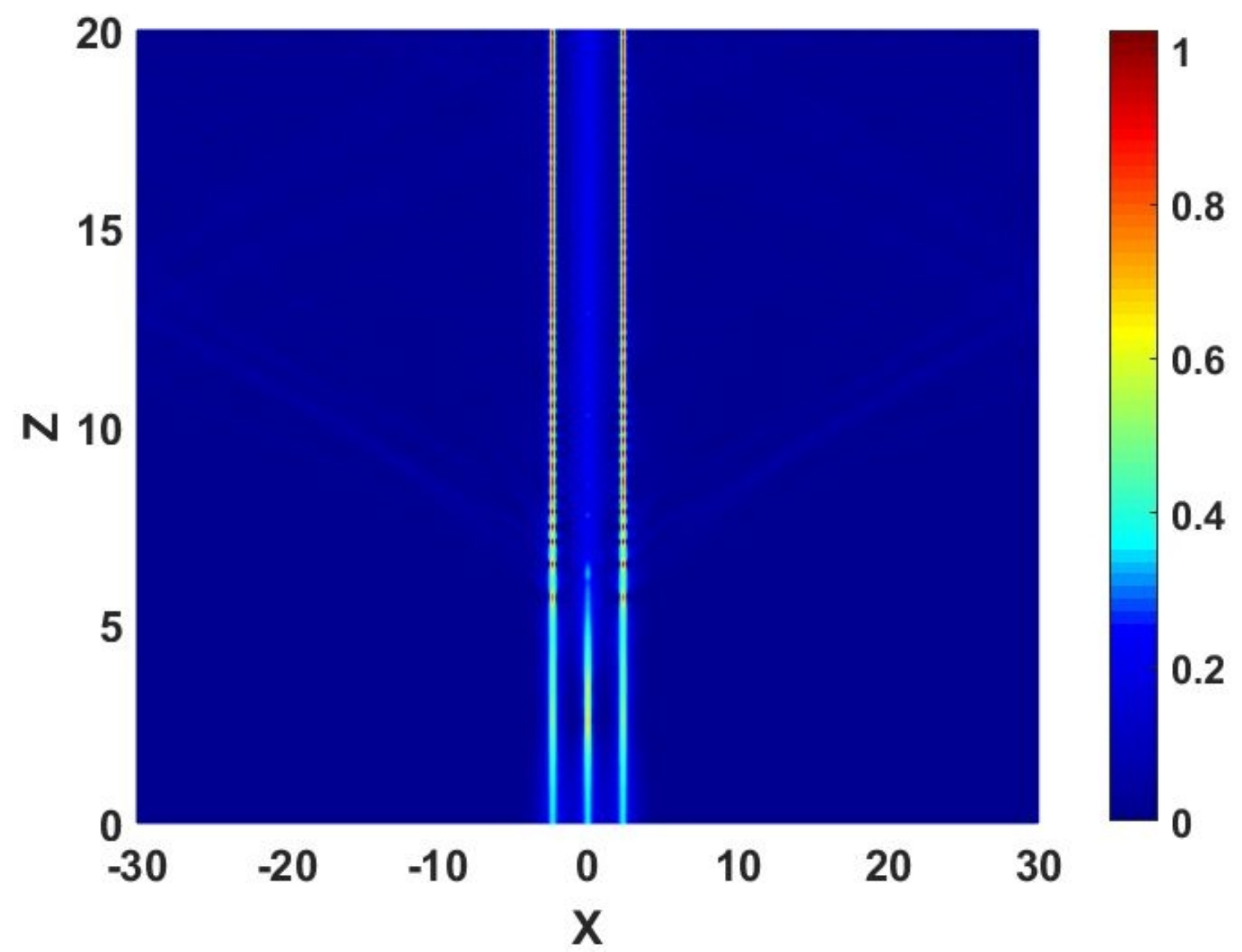}}
\caption{Spontaneous transformation of an unstable in-phase state pinned to
a triple peak, corresponding to the third line in Table \protect\ref{Table1}%
, with $\protect\alpha =0.5$, $\protect\sigma =-1$, $\protect\beta =0$,
propagation constant $k=1.2$ and (a) $\Delta =2.93$, or (b) $\Delta =2.34$.}
\label{fig13}
\end{figure*}

\begin{figure*}[tbp]
\centering\subfloat[]{
	\includegraphics[width=0.49\textwidth]{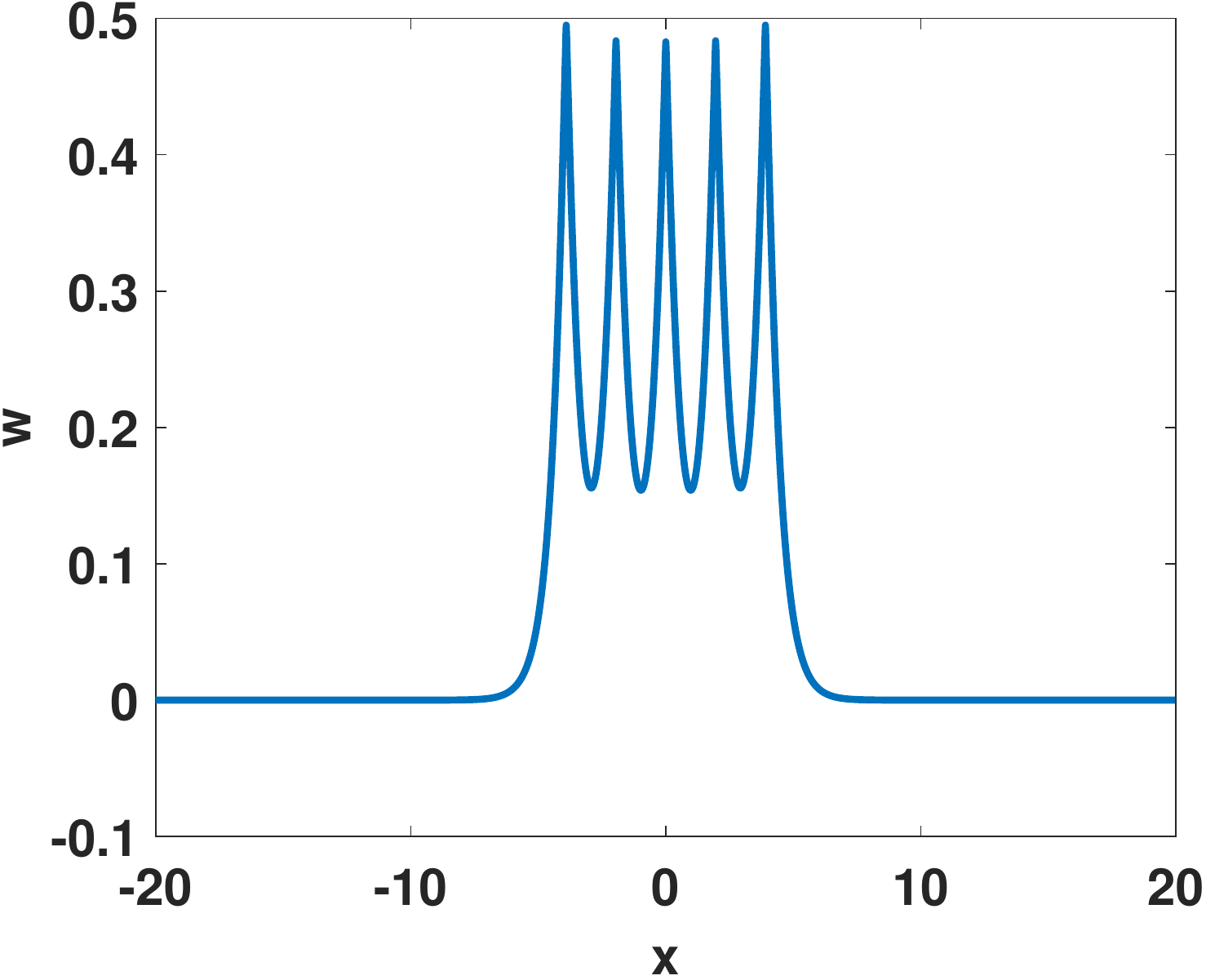}
	}\subfloat[]{
	\includegraphics[width=0.49\textwidth]{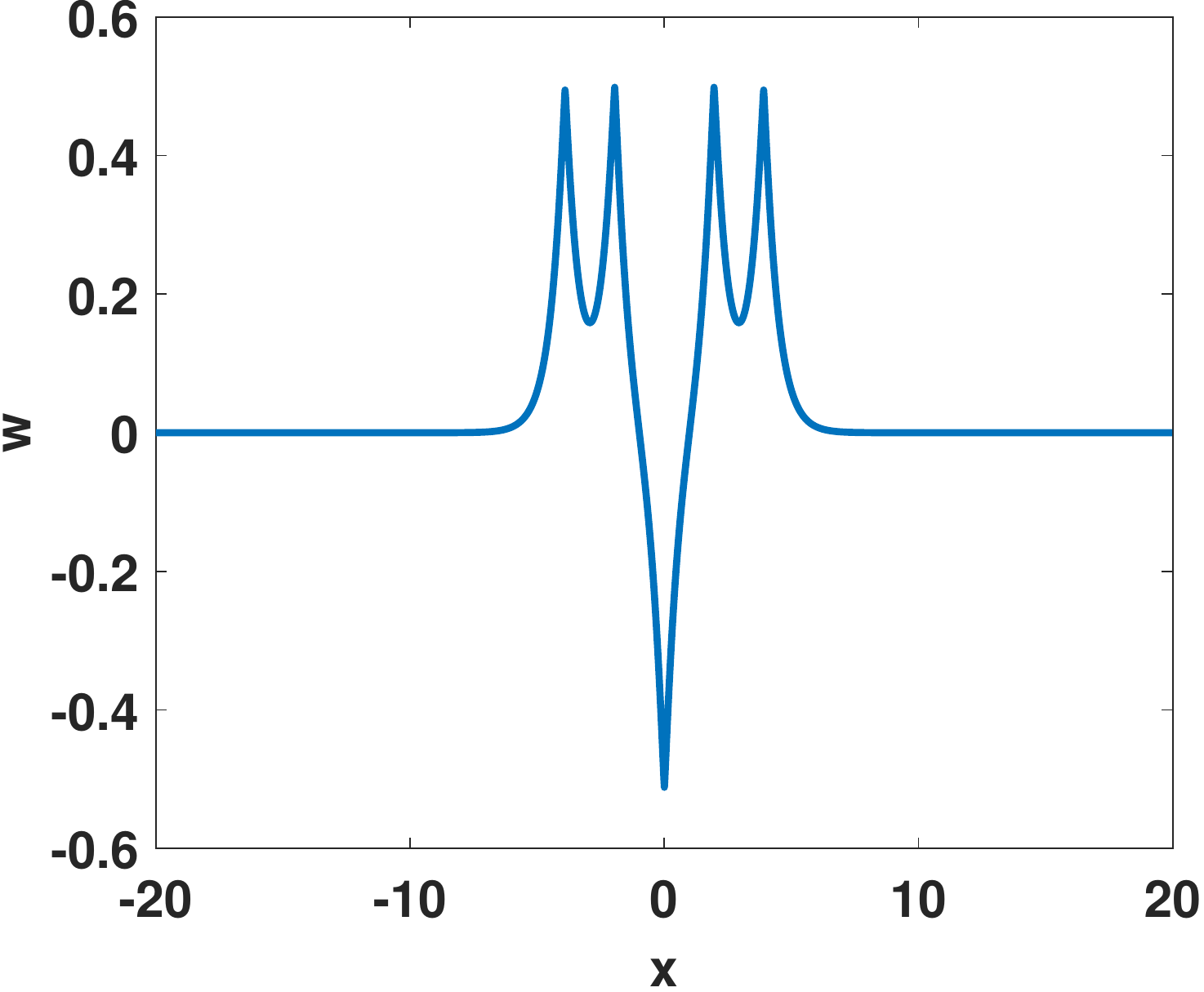}
	}
\caption{Examples of (a) fully in-phase and (b) combined in-phase-twisted
unstable states attached to the set of five singular-nonlinearity-modulation
peaks, which corresponds to the fourth line in Table \protect\ref{Table1}.
The parameters are: $\protect\sigma =-1$, $\protect\alpha =0.5$, $\Delta
=1.95$, $k=2$ and $\protect\beta =0$.}
\label{fig14}
\end{figure*}

As mentioned above, the consideration of the set of five modulation peaks,
which corresponds to the fourth line in Table \ref{Table1}, is relevant too,
as it is a prototype of a periodic lattice of nonlinear defects \cite%
{Barcelona}. In this case, no states with broken symmetry could be produced,
similar to the above-mentioned situation with the triple peak. As concerns
symmetric states, all of them, of both in-phase and twisted types (see
examples in Fig. \ref{fig14}), have been found to be unstable, i.e., the
nonlinear lattice of the singular-modulation peaks, unlike the double and
triple peaks, is not able to support stable states which attach local modes
to all the peaks [unless separation $\Delta $ between adjacent ones is so
large in comparison with widths of the modes pinned to individual peaks that
they cease to interact, or so small that a single narrow mode can cover all
of them, cf. Fig. \ref{fig11}(b)]. In direct simulations (not
shown here in detail), unstable states supported by the five-peak set
transform themselves into ``rarefied" patterns formed by three narrow modes
pinned to the central peak and two edge ones, while intermediate peaks (the
second and fourth ones) remain empty, quite similar to what is shown in Fig. %
\ref{fig13} for unstable states in the case of the triple peak. The emergent
pattern seems stable because individual pinned modes do not interact. This
instability and its development scenario resemble the modulational
instability of extended states pinned to periodic potentials in the NLS
equation with the self-attractive nonlinearity \cite{MI,MI2,Barcelona}.

\subsection{\textbf{Collisions of free solitons with the
nonlinearity-modulation peak}}

Equation Eq. (\ref{eq:u_model_1d}) with the self-focusing uniform
nonlinearity ($\sigma =-1$) obviously supports free NLS solitons far from
the location of the peaks:%
\begin{equation}
u_{\mathrm{sol}}\left( x,z\right) =\frac{P}{2}\exp \left[ ivx+\frac{i}{2}%
\left( \frac{P^{2}}{4}-v^{2}\right) z\right] \mathrm{sech}\left( \frac{P}{2}%
\left( x-vz\right) \right) ,  \label{sol}
\end{equation}%
where $P$ is the soliton's total power, and $v$ is its velocity [in terms of
the spatial optical solitons, it is the tilt of beam in the $\left(
x,z\right) $ plane]. In this case, it is natural to consider collisions of
freely moving solitons with peaks. Here, we limit the consideration to the
basic case of the collision with a single peak, as this case was not studied
previously either.

\begin{figure*}[tbp]
\subfloat[]{
	\includegraphics[width=0.49\textwidth]{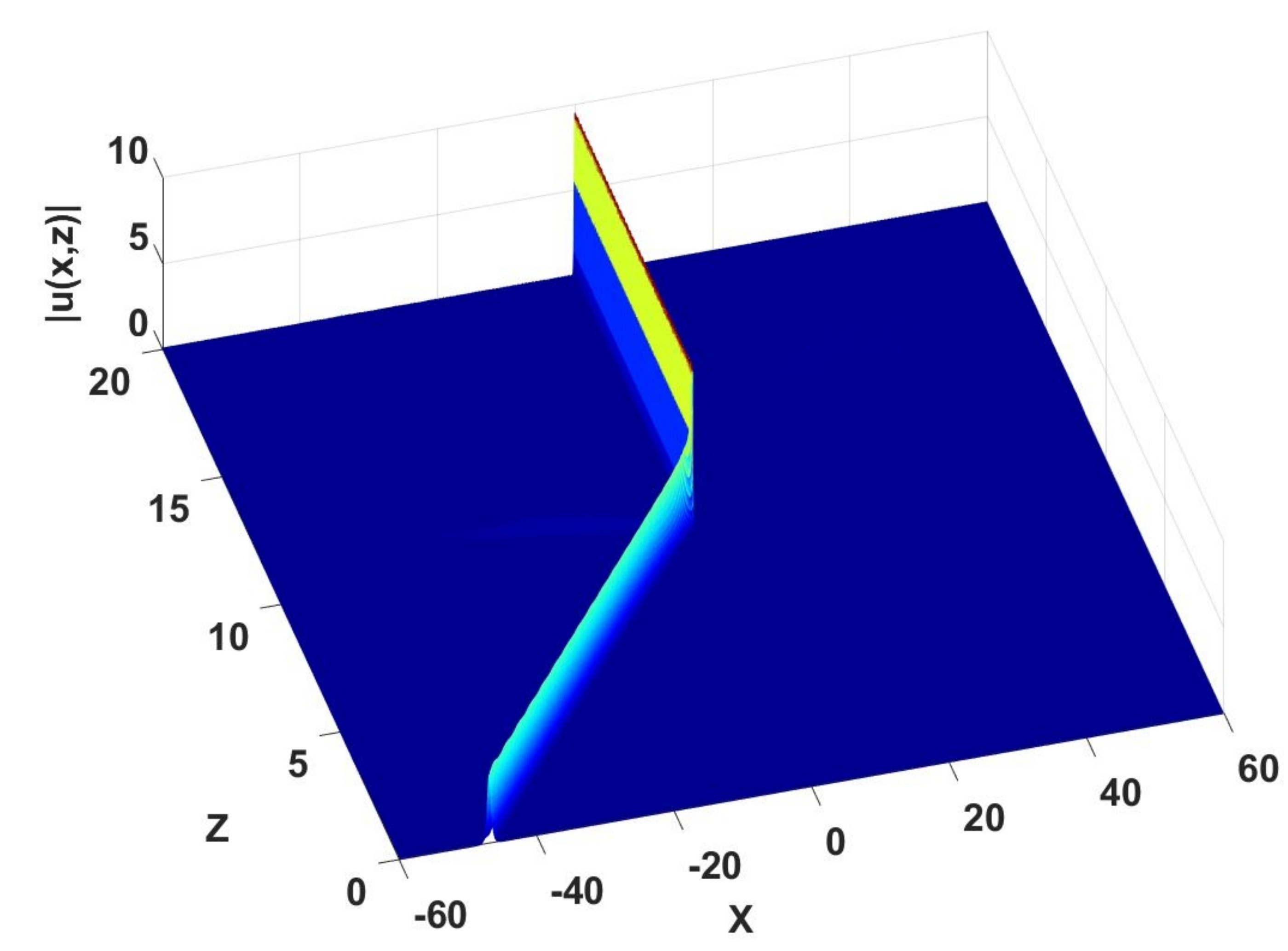}}
\subfloat[]{
	\includegraphics[width=0.49\textwidth]{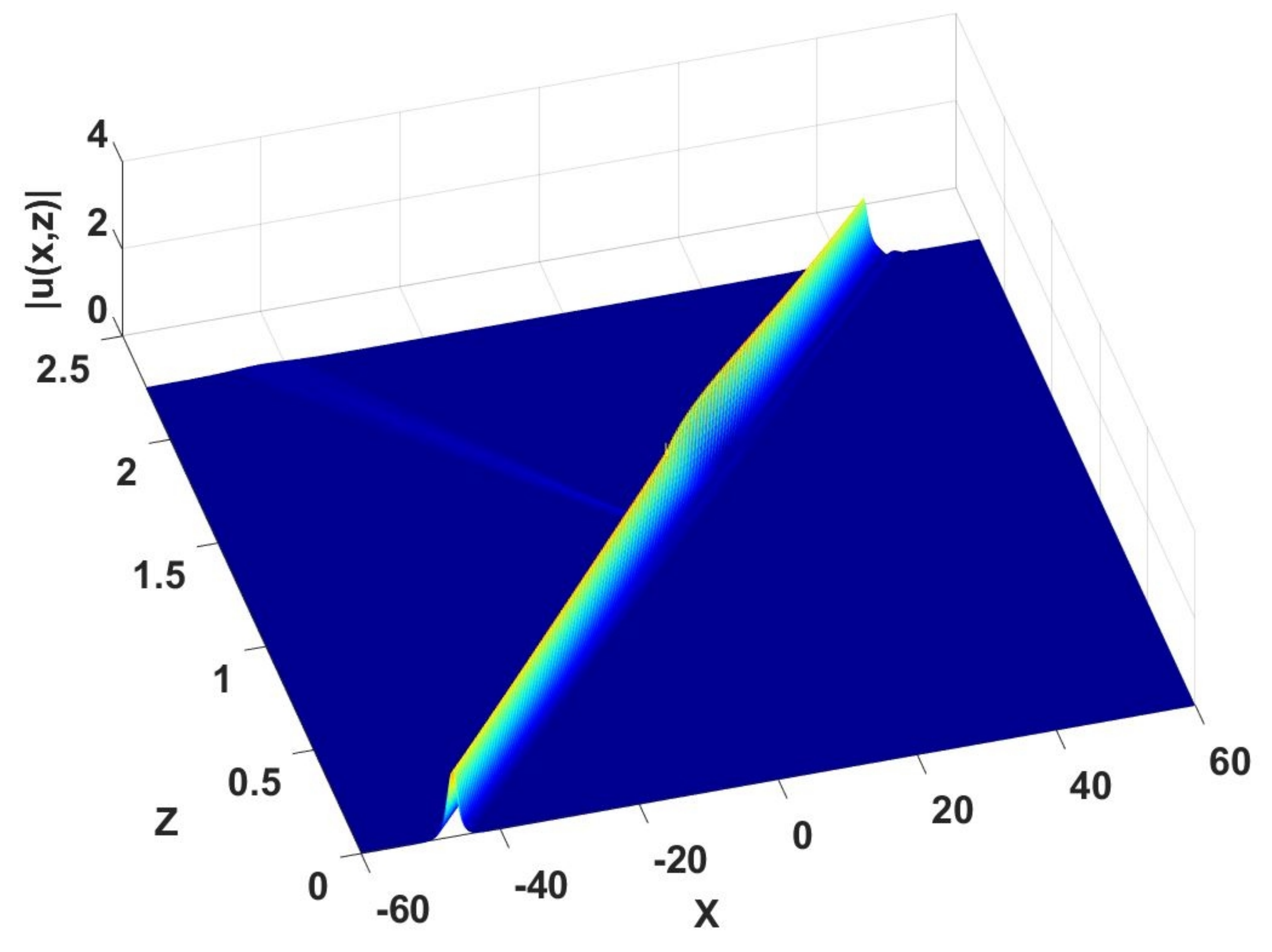}} \newline
\subfloat[]{
	\includegraphics[width=0.49\textwidth]{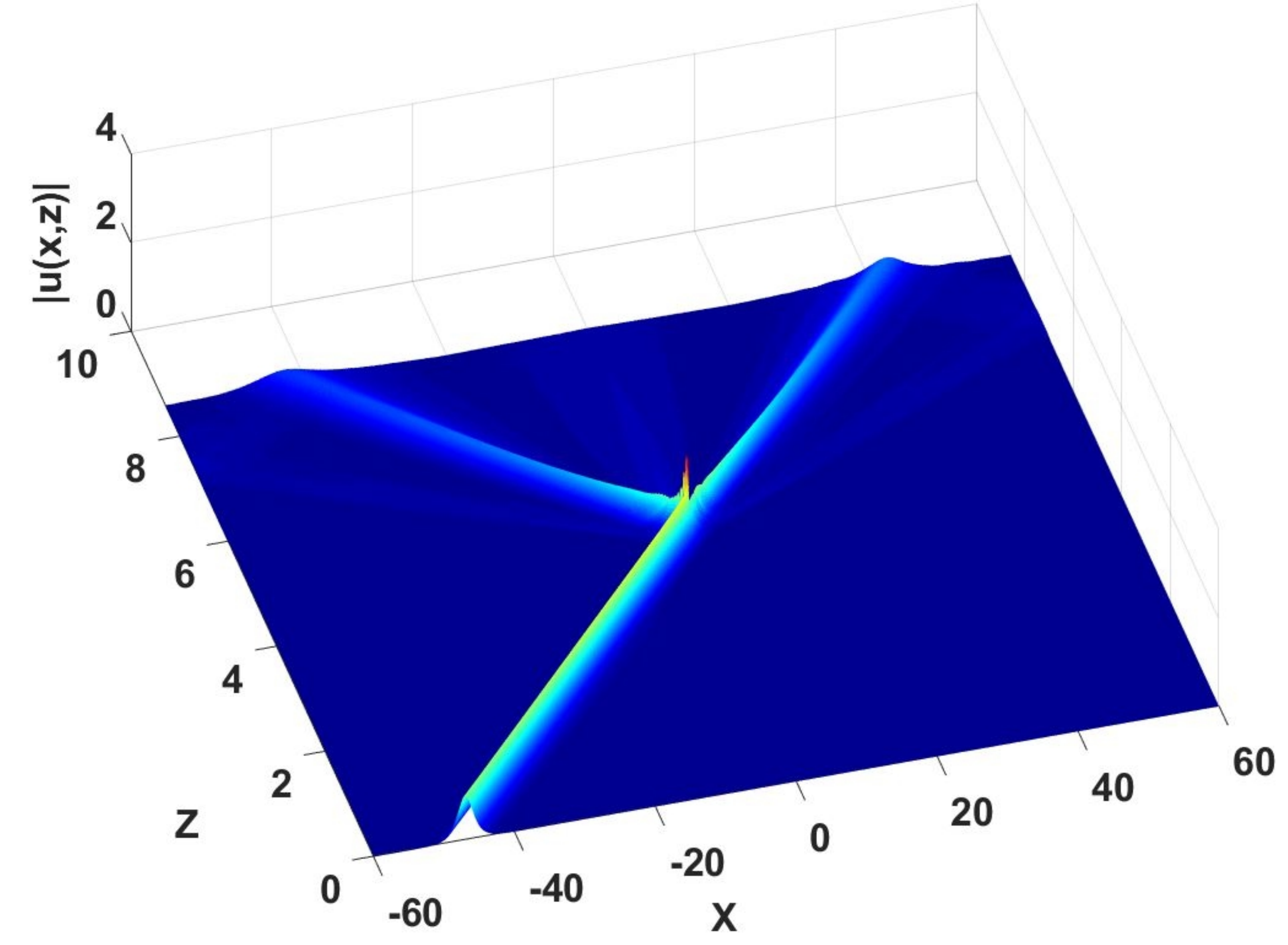}}
\subfloat[]{
	\includegraphics[width=0.49\textwidth]{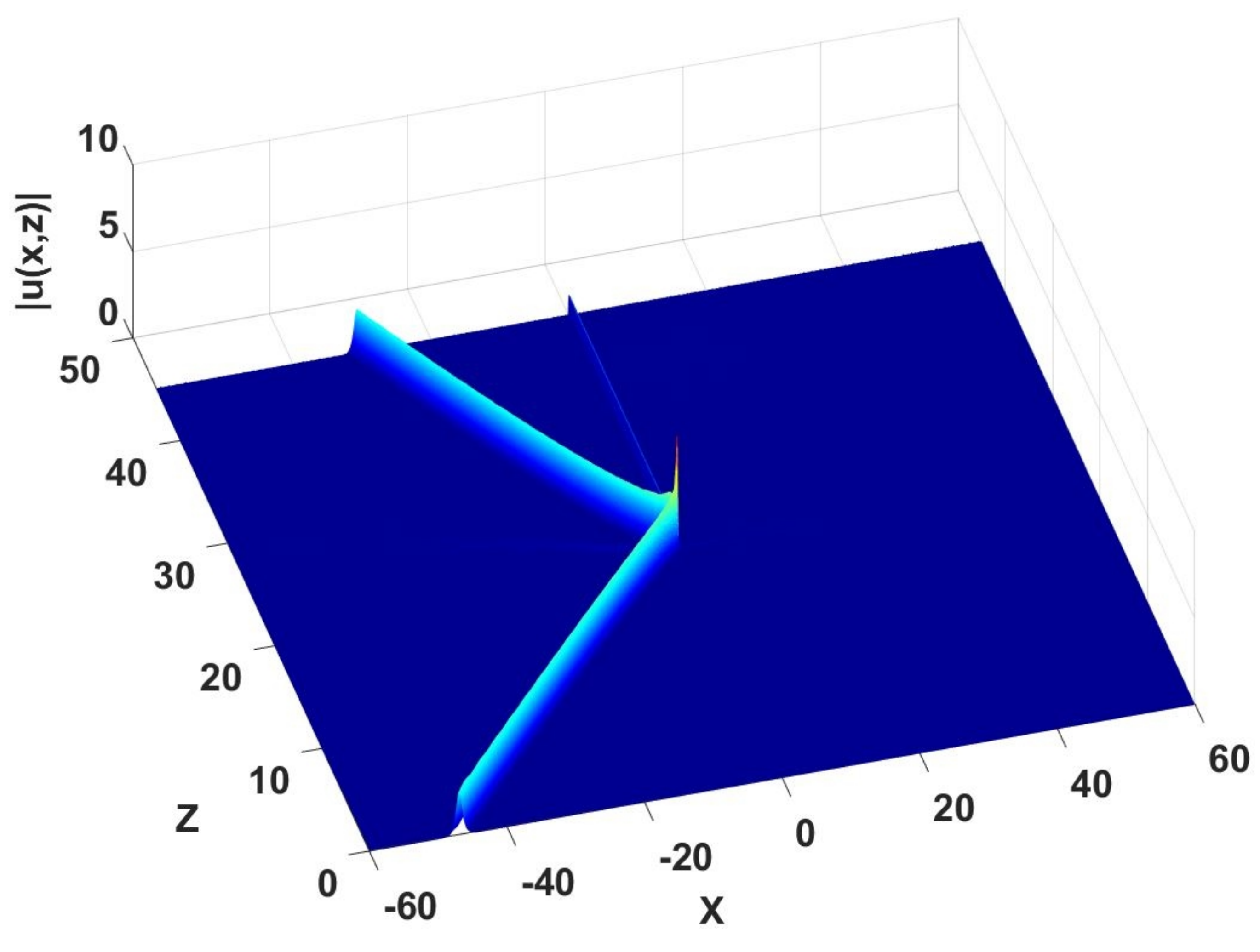}}
\caption{Collisions of an incident free soliton (\protect\ref{sol}) with the
nonlinearity-modulation peak, corresponding to the first line in Table
\protect\ref{Table1} with $\protect\sigma =-1$ and $\protect\beta =0$. (a)
The special case of resonant trapping of the incident soliton with $P=8$ at $%
\protect\alpha =0.6$ and $v=4.5$. (b) Elastic collision of the incident
soliton, with $P=3$ and $v=40$, impinging on the nonlinearity peak with the
singular-modulation power $\protect\alpha =0.3$. (c) Nearly complete
destruction of the incident soliton with $P=2$ at the critical velocity, $v_{%
\mathrm{cr}}=10.5$, for $\protect\alpha =0.5$. (d) The rebound of the
soliton with $P=5$ from the peak with the singular-modulation power $\protect%
\alpha =0.6$ at a small incidence velocity (tilt), $v=2$. }
\label{fig15}
\end{figure*}

\begin{figure*}[tbp]
\centering
\subfloat[]{
		\includegraphics[width=0.32\textwidth]{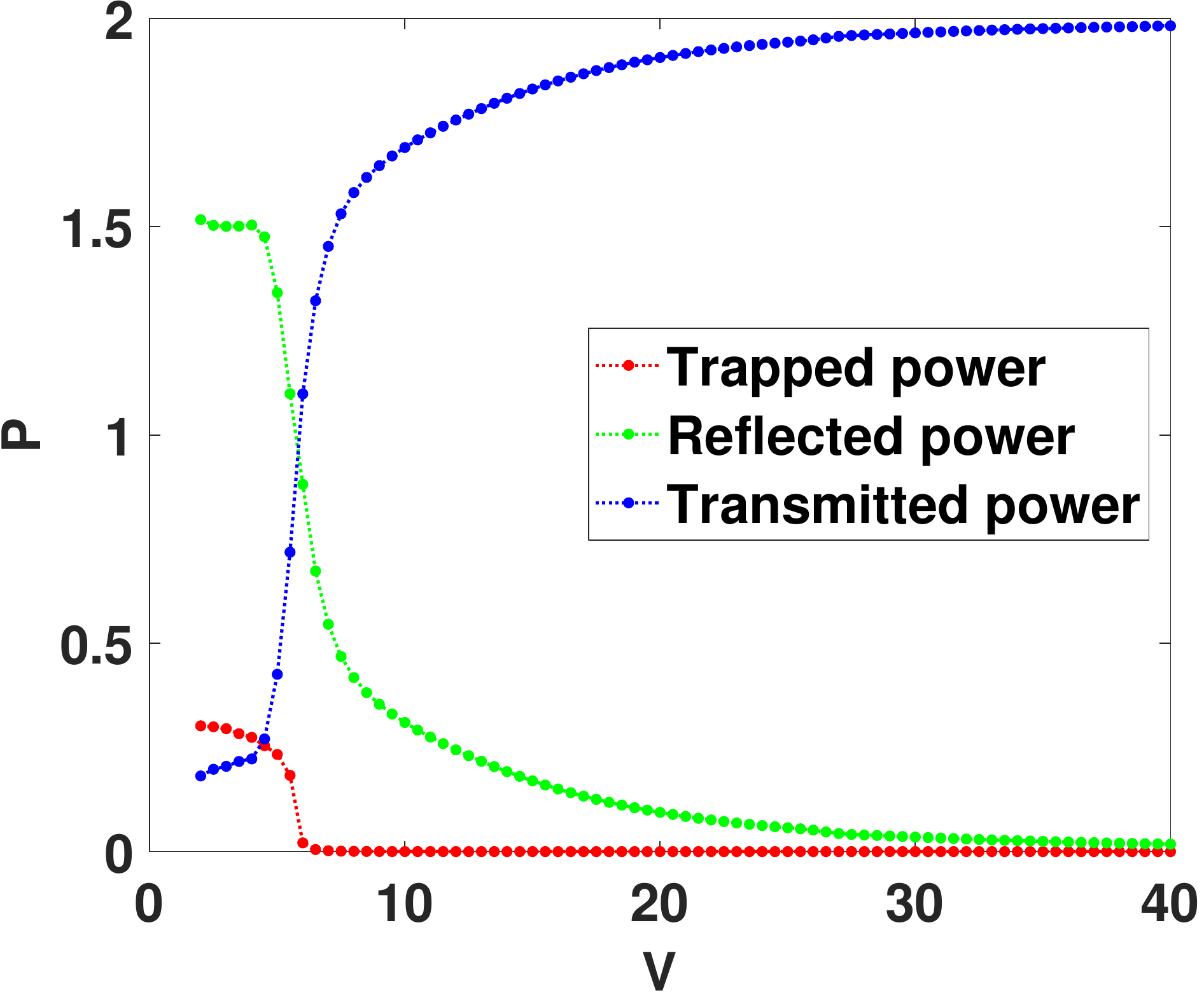}
	} \subfloat[]{
		\includegraphics[width=0.32\textwidth]{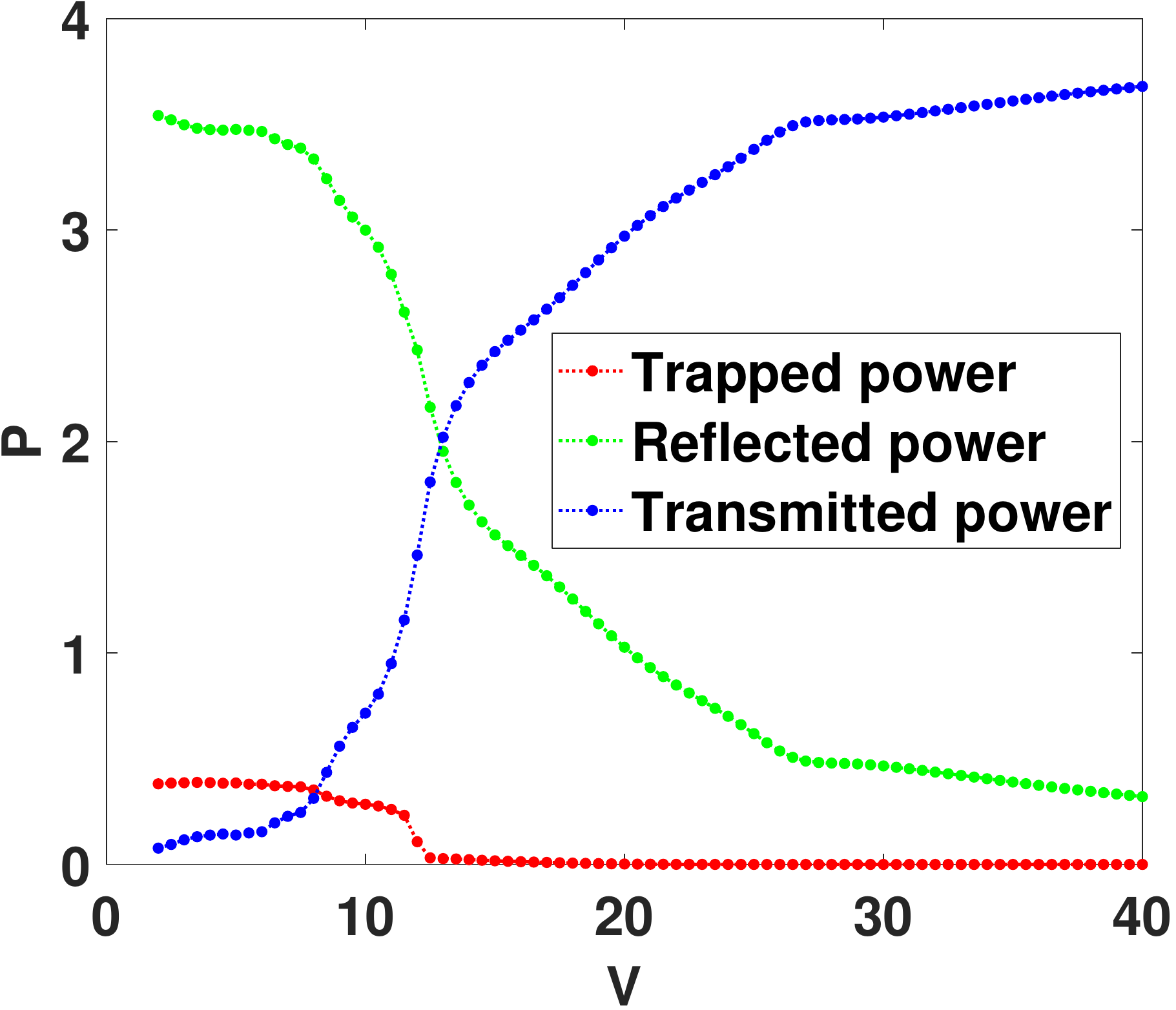}
	}\subfloat[]{
	\includegraphics[width=0.32\textwidth]{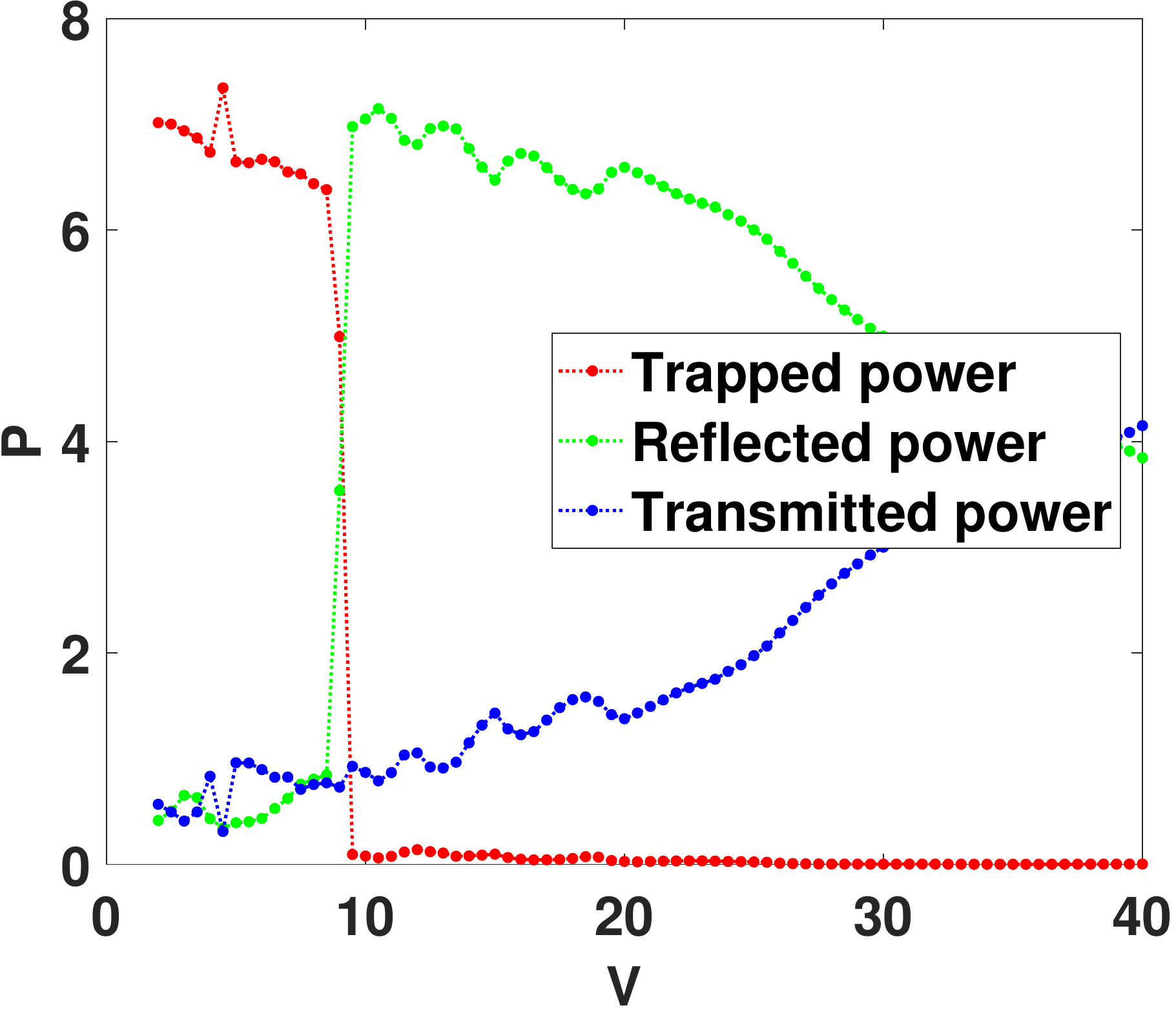}	}
\caption{Shares of the total power of the incident free soliton with (a) $P_{%
\mathrm{sol}}=2$, which corresponds to the propagation constant $k=0.5$ of
the free soliton with zero velocity (tilt) and (b) $P_{\mathrm{sol}}=4$,
which corresponds to propagation constant $k=2$ of the free soliton with
zero velocity (tilt), that are trapped (the red curve) , reflected (the
green curve) and keep the original direction of the motion (``transmitted
power", depicted by the blue curve), versus the incidence velocity, $v$. The
solitons impinge on the nonlinearity peak with $\protect\sigma =-1$, $%
\protect\beta =0$, and $\protect\alpha =0.4$. The respective critical
velocity, at which the soliton is completely destroyed by the collision is $%
v_{\mathrm{cr}}\approx 6$ for the soliton with $P_{\mathrm{sol}}=2$, and $v_{%
\mathrm{cr}}\approx 13$ for the one with $P_{\mathrm{sol}}=4$. Panel (c)
illustrates the special case of the resonant capture of the incident
soliton, an example of which is displayed in Fig. \protect\ref{fig15}(a). In
this case, the parameters are $\protect\sigma =-1$, $\protect\beta =0$, $%
\protect\alpha =0.6$, and $P_{\mathrm{sol}}=8$.}
\label{fig16}
\end{figure*}

An essential result of systematic simulations is that, in most cases, the
incident soliton bounces back from the singular-modulation peak if its
velocity (tilt) is smaller than a certain critical value, $v<v_{\mathrm{cr}}$%
. The soliton is destroyed by the collision into a radiation wave train at $%
v\simeq v_{\mathrm{cr}}$, and, finally, the collision is quasi-elastic at $%
v\gg $ $v_{\mathrm{cr}}$, see typical examples in Figs. \ref{fig15}. The
dependence of the trapped, reflected, and transferred shares of the total
power on the incidence velocity is shown in Fig. \ref{fig16}. The dependence
of $v_{\mathrm{cr}}$ on power $P$ of the impinging soliton for different
values of the singular-modulation power $\alpha $ is shown in Fig. \ref%
{fig17}.

\begin{figure*}[tbp]
\centering
\captionsetup[subfigure]{labelformat=empty}
\subfloat[]{
		\includegraphics[width=0.6\textwidth]{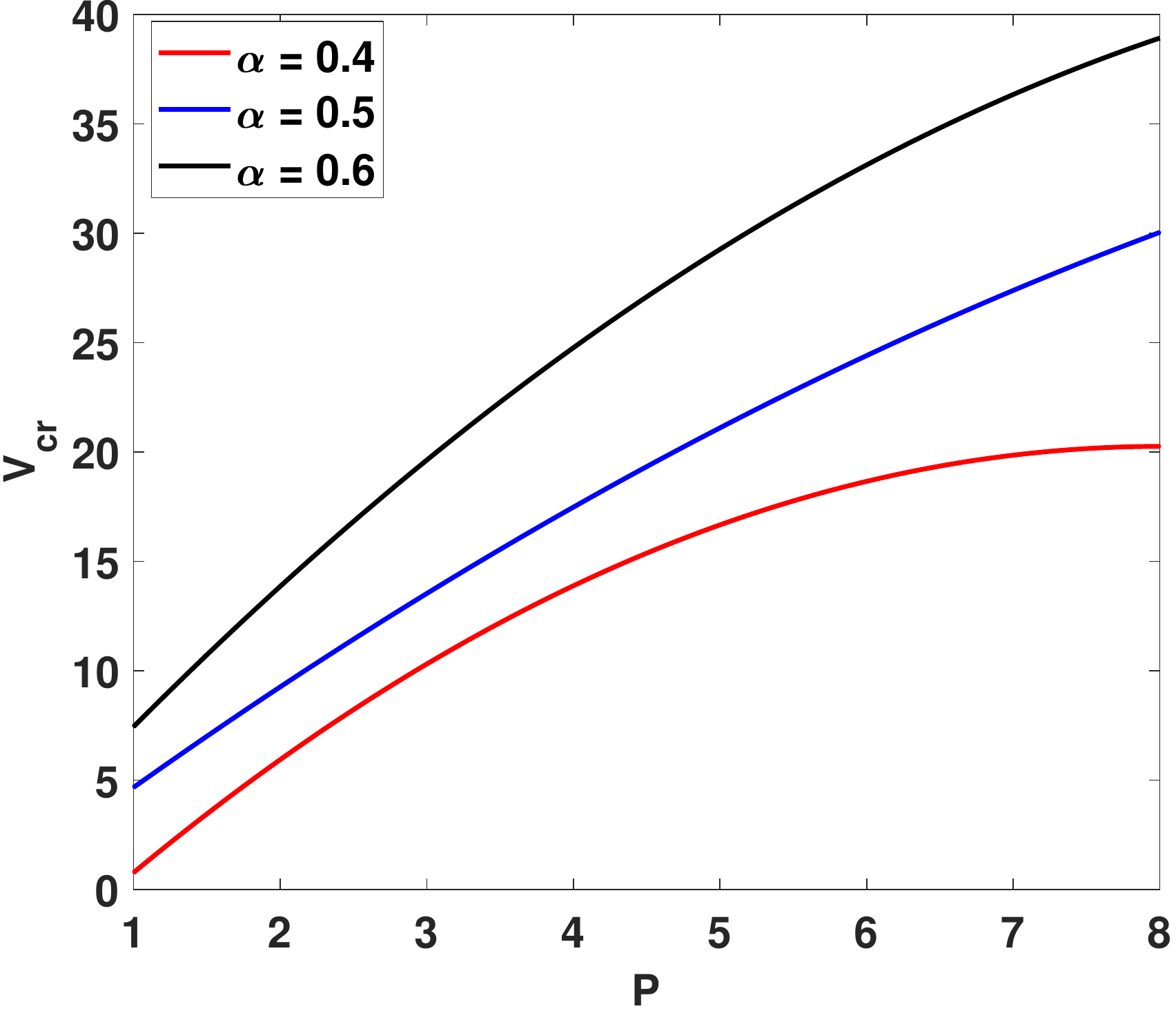}
	}
\caption{$v_{\mathrm{cr}}$ as a function of power $P$ of the moving soliton,
and singular-modulation parameter $\protect\alpha $, at $\protect\sigma =-1$
and $\protect\beta =0$. }
\label{fig17}
\end{figure*}

The fact that the incident soliton readily \emph{bounces back} from the
singular-modulation peak, which represents an \emph{attractive} nonlinear
defect, is remarkable by itself, as rebound is normally expected from
repulsive defects. Previously, rebound of solitons impinging on an
(attractive) \emph{linear }potential well was reported in \cite%
{Brand1,Brand2}.

Capture of the incident soliton by the peak was observed too, but only under
a specific condition, which may be considered as a resonant one: a soliton
with total power $P$ gets captured, as shown in Fig. \ref{fig15}(a), if its
propagation constant in the quiescent state, i.e., $k_{\mathrm{sol}%
}(v=0)=P^{2}/8$, see Eq. (\ref{sol}), is close to the value of $k$ for the
mode with the same value of $P$ trapped by the self-focusing peak, while the
velocity of the incoming soliton is relatively small. This special case is
illustrated by dependences of the of the trapped, reflected, and transferred
shares of the total power on $v$ shown in Fig. \ref{fig16}(c). The resonant
situation can be approximately predicted by equating $k_{\mathrm{sol}}(v=0)$
to the propagation constant predicted, for the same value of $P$, by the VA,
i.e., by the Euler-Lagrange equations following from the effective
Lagrangian (\ref{Leff}) (not shown here in detail).

\section{Conclusions}

We have introduced the 1D model with the cubic self-focusing nonlinearity
locally modulated by a set of singular profiles, while the background
uniform nonlinearity may be focusing or defocusing. The possibility of the
modulation peaks including the linear-potential components is considered
too. Numerical analysis has demonstrated that the setting with a pair of
symmetric peaks readily gives rise to the SSB (spontaneous symmetry
breaking) of the modes pinned to the individual peaks, via the supercritical
bifurcation. The antisymmetric states were found too, turning out to be
unstable and transforming into antisymmetric breathers. Sets of three and
five peaks do not give rise to asymmetric states, rather supporting
symmetric ones, with in-phase or twisted profiles. The twisted states pinned
to the triple peaks have their stability area, while the 
in-phase modes are unstable, except for the strongly overlapping ones. 
The sets of five peaks fails to support any stable nontrivial modes.

\end{document}